\documentclass[draft]{agujournal2019}
\usepackage{url}
\usepackage{lineno}
\usepackage[inline]{trackchanges}
\usepackage{soul}
\usepackage{multirow}
\usepackage{amsmath}

\draftfalse
\journalname{JGR: Earth Surface}

\renewcommand{\baselinestretch}{1.5}\normalsize

\begin{document}

\title{Effect of Discharge on the Nonlinear Bar Growth:\\ A Flume Experiment}
%How Discharge Affects the Growth of Alternate Bars: A Flume Experiment
%Origin: Discharge Dependence of the Linear Growth Rate of Alternate Bars
%Response, Sensitivity

\authors{S. Seki\affil{1}, D. Moteki\affil{2}, H. Yasuda\affil{2}}
\affiliation{1}{Graduate School of Science and Technology, Niigata University, Niigata, Japan}
\affiliation{2}{Institute of Industrial Science, The University of Tokyo, Tokyo, Japan}
\correspondingauthor{Hiroyasu Yasuda}{hiroyasu@iis.u-tokyo.ac.jp}

\begin{keypoints}
\item Repeated bed measurements over time suggest that alternate-bar height tends to grow faster at lower discharge on a bedload-based timescale  %138/140
\item Growth rate of bar height depends less on discharge than linear growth rate predicted by stability theory for amplitude of a single bar mode  %140/140
\item During bar growth, lower discharge strengthens flow deflection, enlarges non-transport zones, slows bar migration, and increases bar height %139/140
\end{keypoints}
%/140characters

\begin{abstract}
Alternate bars are ubiquitous bedforms in alluvial rivers, and excessive bar-height growth can increase flood risk and disrupt ecosystems.
Predicting bar-height variability requires quantifying the dependence of bar growth rate on water discharge, yet this has rarely been achieved because conventional flume measurements interrupt the flow and cannot continuously capture bar evolution.
Here we conducted laboratory experiments under four discharges and four channel slopes (16 conditions) and continuously mapped bed topography using Stream Tomography, a nonintrusive, flow-through measurement technique.
Bar height exhibits sigmoidal growth and is well described by the Landau equation, allowing robust estimation of the growth rate and equilibrium height.
The inferred dimensionless growth rate tends to increase as discharge decreases, but its sensitivity is weaker than that predicted by stability theory for a single alternate-bar mode.
This discrepancy becomes more pronounced at lower discharge, indicating limitations of linearization and the influence of interactions among multiple bar modes.
Furthermore, numerical flow simulations over the measured bed topography reveal that, during bar development, lower-discharge conditions produce a relatively larger bar height-to-depth ratio, making the flow more prone to deflection.
In such cases, sediment transport is restricted downstream of depositional areas, thereby limiting bar migration. 
Consequently, flow and sediment transport become concentrated in the scour zones, enhancing local scour and promoting the growth of bar height.
These results provide benchmark constraints for bar stability theories and help improve predictions of how bar height responds to changes in discharge.
\end{abstract}
%235/249words

\section*{Plain Language Summary}
Most rivers develop sandbars that alternate along opposite banks.
As sandbars grow taller, they change aquatic habitats and increase flood hazards through bank erosion and deep scour.
How rapidly the bar height grows depends on river flow, but measuring it has been challenging because traditional experiments require stopping the flow.
We conducted laboratory experiments that continuously mapped the riverbed under flowing water across different discharges.
This allowed us to track the bar development from initial formation to a stable final state.
In each experiment, bar height increased rapidly at first and then gradually leveled off, following a sigmoidal curve.
We found that bars grow faster at lower discharges using a timescale based on sediment transport, although their sensitivity to discharge is weaker than existing theories predict.
Hydraulic calculations over the measured bed explain why low discharges promote bar growth: during bar development, gravity bends the flow toward the deep channels near the banks.
At the same time, sediment transport is reduced downstream of depositional areas, so bars migrate slowly.
This concentrates flow and sediment transport in the scour zone, helping bars grow higher.
Our results provide new experimental evidence to test and refine widely used theories of bed evolution.
%199/199words

%========================================================1============================================================
\newpage
\section{Introduction}
In natural rivers, periodic bedforms spanning a wide range of scales---classified as micro-scale, meso-scale, and macro-scale depending on geomorphic setting---are observed \cite{Seminara2010, Dey2020}.
In alluvial rivers, meso-scale bars commonly form in reaches spanning from alluvial fans to natural levee zones.
Bars are three-dimensional morphologies with wavelengths several times the channel width and amplitudes (heights) on the order of flow depth.
Excessive growth of bar height deflects the flow, leading to local scour and bank erosion \cite<e.g.,>{Fujita1982, Visconti2010, Nagata2014, Inoue2025}.
In addition, as the frequency of inundation over depositional areas decreases, vegetation can establish and grow easily.
This not only reduces channel conveyance and obstructs flood flows, but also alters the original river ecosystem \cite<e.g.,>{Gurnell2014, Hauer2016, Hu2024, Holusova2025}.
It therefore remains a critical challenge in both science and engineering to elucidate the mechanisms controlling bar-height development and to predict its variability. 

Changes in river morphology, including bars, are governed by channel geometry (width and bed slope), bed material properties (composition and grain size), and water discharge.
Among these factors, channel geometry and bed material can be regarded as approximately constant over time unless large floods induce substantial bank erosion and sediment redistribution, or unless anthropogenic interventions occur.
In contrast, discharge is an external forcing that influences river morphology through its effects on flow velocity and flow depth, and it can vary markedly even within a single year.
Moreover, flood magnitude and frequency are directly affected by the impacts of climate change that have become increasingly apparent in recent decades \cite<e.g.,>{Hirabayashi2013, Arnell2016, Tanoue2016, Bloschl2019, Slater2021}.
Thus, discharge can be identified as a key factor influencing the variability of bar height \cite{Redolfi2020, Nelson2018}.

Bar stability analysis is one useful approach for predicting how channels respond to changes in discharge.
In linear stability analysis, the governing equations for shallow-water flow and bed evolution are first linearized, and small perturbations are then introduced.
When the linear growth rate (i.e., the temporal coefficient of the perturbation) is zero, the bed is considered neutrally stable.
The corresponding width-to-depth ratio then provides the critical condition for bar formation.
Previous studies have confirmed that this criterion works well as an indicator of bar formation \cite{Kuroki1984, Colombini1987, Redolfi2021}.
These analyses were later extended to more complex conditions relevant to natural rivers, including grain sorting \cite{Lanzoni1999}, secondary-flow effects \cite{Iwasaki2016}, and vegetation \cite{Barenbold2016}. 
In parallel, \citeA{Tubino1999} provided a comprehensive review of free-bar dynamics, covering three-dimensional formulations and the roles of suspended load, grain sorting, and channel-width variations in bar instability.
However, within the linear framework, perturbations are assumed to grow exponentially with time, which makes it difficult to accurately predict bar height that reaches a saturated value even under steady discharge \cite{Fujita1985}.
To address this limitation, \citeA{Colombini1987} developed a weakly nonlinear stability theory for bars.
The theory assumes that the evolution of bar amplitude toward a finite value can be described by a Landau equation, which yields a self-limiting temporal trajectory.
The Landau equation is governed by two coefficients---the linear growth rate and the equilibrium amplitude.
Accurate inference of these coefficients enables quantitative prediction of bar height, which varies with discharge.

Of the two coefficients in the Landau equation, the equilibrium amplitude determines the maximum bar height attainable under a given set of conditions.
Equilibrium bar height has been investigated extensively because it can be obtained relatively easily from final bed topographies.
From the 1950s to the 1980s, flume experiments on bars were actively conducted, providing a large body of fundamental data \cite<e.g.,>{Kinoshita1961, Ikeda1983, Fujita1982, Fujita1985}.
Subsequent studies used these experimental data to develop predictive equations for bar height through dimensional analysis \cite{Jaeggi1984, Ikeda1984, Cheng2019}.
Weakly nonlinear theory has also been used to predict the finite amplitude of bars, and comparisons with available experimental data have shown reasonable agreement \cite{Colombini1987, Ali2021}.
As with linear theory, this line of research has naturally evolved toward incorporating more complex assumptions. 
For example, \citeA{Tubino1991} analyzed bar development under unsteady flow conditions designed to mimic natural flood events, and the validity of this approach was later examined using the field observations of \citeA{Welford1994}. 
More recent studies have continued to investigate bar dynamics under unsteady flow conditions \cite{Hall2004, Izumi2021, Izumi2022}.
Furthermore, \citeA{Bertagni2018} applied a nonlinear analysis that accounts for suspended load effects to the Mississippi River. 
In stability analyses of gravel-bed rivers, sediment transport is generally assumed to be bedload \cite{Colombini1987}. 
\citeA{Bertagni2018} showed that the discrepancy between theory and observation can be reduced relative to analyses that consider bedload only.
Thus, observational data enable the applicability of theory to be tested and refined, thereby improving predictive accuracy for bar height.

The linear growth rate, the other coefficient in the Landau equation, controls the slope of the temporal evolution of bar amplitude.
The relationship between the linear growth rate and discharge can be inferred from linear stability theory \cite{Carlin2021, Izumi2021, Izumi2022}.
Quantifying the growth rate, however, remains challenging in both natural rivers and laboratory flumes.
This is largely because reliable estimation of growth rate requires densely sampled time-series data, yet many conventional approaches---even in relatively controllable flume experiments---interrupt the flow at each stage of bar development \cite<e.g.,>{Fujita1985}, thereby limiting the temporal resolution of bed-topography measurements.
Moreover, whereas stability analysis generally focuses on the linear growth rate associated with the most amplified wavelength of the alternate-bar mode, actual alternate bars are expected to be represented not only by the alternate-bar mode itself but also by its superposition with higher-wavenumber modes \cite{Hasegawa1982, Garcia1993, Redolfi2020}.
The measurement limitation has made it difficult to assess how well the maximum linear growth rate of the amplitude of alternate-bar mode in stability theory approximates the growth rate of alternate-bar height.
To overcome this limitation, we developed Stream Tomography (ST) \cite{ST}.
ST can obtain planimetric bed topography beneath the water surface by repeatedly moving a laser-equipped cart back and forth over the flume.
In other words, it enables the entire process of bar development, from an initially flat bed to the equilibrium state, to be captured without interrupting the flow \cite{Ishihara2022, Moteki2023, Seki2023}.
Moreover, its temporal and spatial resolutions are 1 min and 1 cm$^2$, respectively, which are sufficiently high compared with those of conventional experimental methods.
In this study, we accurately quantify the growth rate of bar height by ST and clarify its relationship with discharge.

It is also important to understand the physical mechanisms that control how the bar growth rate varies with discharge.
We propose the following hypothesis and test it by combining high-frequency ST measurements of bed topography with hydraulic calculations.
Specifically, under low-discharge conditions, during the stage of bar development, the bar height-to-depth ratio is relatively large, so the flow is strongly deflected by the bar topography.
At the same time, because the main flow does not pass through the region immediately downstream of the depositional area, non-sediment-transport zones develop and expand, thereby suppressing bar migration.
Under these conditions, flow and sediment transport become concentrated in the scour zone, which in turn promotes the growth of bar height.
Conversely, under high-discharge conditions, during the stage of bar development, the height-to-depth ratio is relatively small, so the flow is straighter.
At the same time, because sediment transport occurs over most of the bar surface, the bars migrate downstream.
Under these conditions, scour location is less likely to remain fixed, which in turn suppresses the growth of bar height.

In this study, we focus on water discharge as a key factor controlling variations in bar height, with the objectives of (i) quantifying the dependence of the growth rate of alternate-bar height on discharge and (ii) explaining the underlying physical mechanisms.
First, we quantify the development process of bar height using a continuous bed-topography measurement technique in laboratory flume experiments.
We then fit a Landau equation to the resulting time series to infer the growth rate with high accuracy and to clarify its relationship with discharge.
We further compare the experimentally inferred growth rates of bar height with values of the amplitude of a single alternate-bar mode predicted by linear stability analysis and discuss the causes of any discrepancies.
Second, using fixed-bed hydraulic calculations with the measured bed topography prescribed, we evaluate how flow deflection and the extent of non-sediment-transport zones vary with discharge.
On this basis, we provide a mechanistic interpretation of the relationship between the bar growth rate and discharge.
This paper is organized as follows:
Section 2 describes the experimental and analytical methods.
Section 3 presents the experimental results.
Section 4 presents the hydraulic calculation results.
Section 5 discusses the experimental and hydraulic calculation results.
Section 6 summarizes the conclusions.
%========================================================1============================================================

%========================================================2============================================================
\newpage
\section{Methods}
\subsection{Experimental Setup}
\begin{figure}[t]
 \centering
   \includegraphics[width=1.0\textwidth]{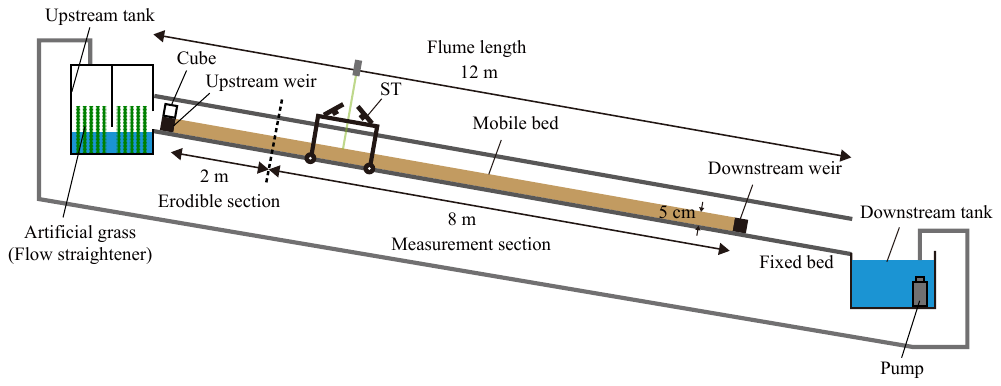}
 \caption{
 Longitudinal view of the experimental flume.
 }
 \label{flume}
\end{figure}

The channel used for the experiment is shown in Figure \ref{flume}.
The flume has a rectangular cross-section (width $B$ = 45\ cm), and it is made of fiber-reinforced plastic.
The mobile-bed section was 10 m\ long and consisted of a 5\ cm-thick layer of uniformly spread silica sand with an average grain size $d$ = 0.755\ mm, which was leveled to be flat before the start of each run.
The sediment layer was sufficiently thicker than the maximum scour depth observed in the experiments, preventing exposure of the fixed bed.
The channel was constructed with weirs 5\ cm high, 5\ cm wide, and 45\ cm long at the upstream and downstream ends of the mobile-bed section. 
Because the flow depth was relatively shallow, approximately 1--2\ cm, the weirs were installed to maintain stable shallow flow over the relatively thick sediment layer.
The upstream weir maintained the water surface near the upstream end of the mobile-bed section, whereas the downstream weir controlled the downstream water level.
%The length of the mobile-bed section was 10\ m.
%The channel was constructed with weirs 5\ cm high, 5\ cm wide, and 45\ cm long at the upstream and downstream ends of the mobile-bed section. 
%As the initial condition, the channel was given a flat bed of uniformly spread silica sand with an average grain size $d$ = 0.755\ mm.
No sediment was supplied during the run to minimize human interference.
Instead, the upstream 2\ m of the 10\ m mobile-bed reach was designated as a sediment-supply section, and measurements were restricted to the downstream 8\ m.
To shorten the development length required for bar formation, we installed a rubber cube with an edge length of 5\ cm on the left bank at the upstream end as a perturbation source \cite<e.g.,>{Crosato2012, Moteki2023}.

The hydraulic conditions are summarized in Table \ref{condition}.
The water discharge $Q$ was set to four values, 3.2, 2.7, 2.2, and 1.7\ L/s.
These discharges were kept steady, as confirmed using an electromagnetic flow meter.
Flow circulation was maintained using a pump.
$Q_\mathrm{c}$ denotes the critical discharge for bar formation, calculated by the method described later.
If the imposed discharge exceeds this value, bar formation is theoretically suppressed, so $Q$ was set below this threshold.
Four channel slopes $I$ = 1/91, 1/111, 1/143, and 1/200 were employed to examine the possible dependence of bar height response to discharge on channel slope.
Combining these four discharges with the four slopes resulted in a total of 16 hydraulic conditions.
The hydraulic conditions were designed, with reference to the classification diagram by \citeA{Kuroki1984}, so that the combination of $BI^{0.2}/h_0$ and the Shields number $\tau_{*0}$ would fall within the alternate bar domain, where $h_0$ denotes the uniform flow depth.

It is desirable to adopt a common timescale because the hydraulic conditions considered span different combinations of discharge and channel slope.
A characteristic timescale for bed evolution is provided by the Exner equation; experimental time was nondimensionalized using this timescale \cite<e.g.,>{Redolfi2021}.
The experiments were continued until the dimensionless time $t^*$ defined below reached approximately 30:
%%%%%%%%%%%%%%%%%%%%%%%%%%%%%%%%%%%
\begin{equation}
t^* = t \cdot \frac{q_{\mathrm{s}0}}{(1-p)h_0B/2},
\label{t*}
\end{equation}
%%%%%%%%%%%%%%%%%%%%%%%%%%%%%%%%%%%
\begin{equation}
q_{\mathrm{s}0} = 8(\tau_{*0}-\tau_{*\mathrm{c}})^{3/2}\sqrt{sgd^3},
\label{MPM}
\end{equation}
%%%%%%%%%%%%%%%%%%%%%%%%%%%%%%%%%%%
where $t$ is the experimental time, $q_{\mathrm{s}0}$ is the bedload transport at uniform flow by \citeA{mpm}, $p$ = 0.4 is the bed porosity, $\tau_{*\mathrm{c}}$ = 0.034 is critical Shields number by \citeA{Iwagaki1956}, $s$ = 1.65 is the specific gravity of the sediments in the water, $g$ = 9.81 \ m/s$^2$ is the gravitational acceleration.

We used Stream Tomography (ST) \cite{ST} to capture the temporal evolution of the bed topography.
As shown in Figure \ref{flume}, this method measures the bed topography while maintaining the flow, because a cart carrying a laser-sheet light source and a camera moves back and forth over the flume.
The measurement interval was 1 min.
We conducted at least five runs under each condition to account for the inherent variability of bar formation.
In this paper, we refer to each run as a “Replicate”.

%=======================================================Figs==========================================================

\renewcommand{\baselinestretch}{1}\normalsize
\begin{table}[t]
\caption{Hydraulic conditions.}
\centering
\begin{tabular}{ccccccccccccc}
\hline
Case & $I$ (-) & $Q_{\mathrm c}$ (L/s) & $Q$ (L/s) & $BI^{0.2}/h_0$ (-) & $\mathrm{Fr}^a$ (-) & ${\tau_{*0}}^b$ (-) & $h_0$ (cm) & ${u_0}^c$ (cm/s) & ${t_{\mathrm{e}}}^d$ (min)\\
\hline
1 & \multirow{4}{*}{$1/91$} & \multirow{4}{*}{4.84} & 3.2 & 11.8 & 1.18 & 0.136 & 1.54 & 46.1 & 50 \\
2 &  &  & 2.7 & 13.1 & 1.16 & 0.123 & 1.39 & 43.0 & 55 \\
3 &  &  & 2.2 & 14.8 & 1.14 & 0.109 & 1.23 & 39.6 & 60 \\
4 &  &  & 1.7 & 17.3 & 1.11 & 0.093 & 1.06 & 35.8 & 75 \\
\hline
5 & \multirow{4}{*}{$1/111$} & \multirow{4}{*}{4.36} & 3.2 & 10.7 & 1.08 & 0.118 & 1.64 & 43.4 & 70 \\
6 &  &  & 2.7 & 11.8 & 1.06 & 0.107 & 1.48 & 40.5 & 80 \\
7 &  &  & 2.2 & 13.4 & 1.04 & 0.095 & 1.31 & 37.3 & 90 \\
8 &  &  & 1.7 & 15.6 & 1.02 & 0.081 & 1.12 & 33.7 & 115 \\
\hline
9 & \multirow{4}{*}{$1/143$} & \multirow{4}{*}{3.91} & 3.2 & 9.4 & 0.97 & 0.099 & 1.77 & 40.2 & 110 \\
10 &  &  & 2.7 & 10.4 & 0.95 & 0.090 & 1.60 & 37.6 & 125 \\
11 &  &  & 2.2 & 11.8 & 0.93 & 0.079 & 1.41 & 34.6 & 150 \\
12 &  &  & 1.7 & 13.8 & 0.91 & 0.068 & 1.21 & 31.2 & 200 \\
\hline
13 & \multirow{4}{*}{$1/200$} & \multirow{4}{*}{3.52} & 3.2 &  8.0 & 0.83 & 0.079 & 1.96 & 36.4 & 215 \\
14 &  &  & 2.7 & 8.8 & 0.82 & 0.071 & 1.77 & 34.0 & 260 \\
15 &  &  & 2.2 & 10.0 & 0.80 & 0.063 & 1.56 & 31.3 & 340 \\
16 &  &  & 1.7 & 11.7 & 0.78 & 0.054 & 1.34 & 28.2 & 510 \\
\hline
\multicolumn{11}{p{16cm}}{
$^a$Froude number $\mathrm{Fr}$ = $u_0/\sqrt{gh_0}$. 
$^b$Shields number $\tau_{*0}$ = ${h_0 I}/{sd}$.
$^c$Manning velocity equation $u_0$ = ${n_s}^{-1} {h_0}^{2/3} I^{1/2}$, $n_s$ = 0.014\ (m$^{-1/3}\cdot$s). 
$^d$Time when the run was terminated. Some replicates ended earlier than this time.
}
\label{condition}
\end{tabular}
\end{table}
\renewcommand{\baselinestretch}{1.4}\normalsize
%=======================================================Figs==========================================================

\subsection{Definition of the Alternate-Bar Height}
\begin{figure}[t]
 \centering
   \includegraphics[width=0.6\textwidth]{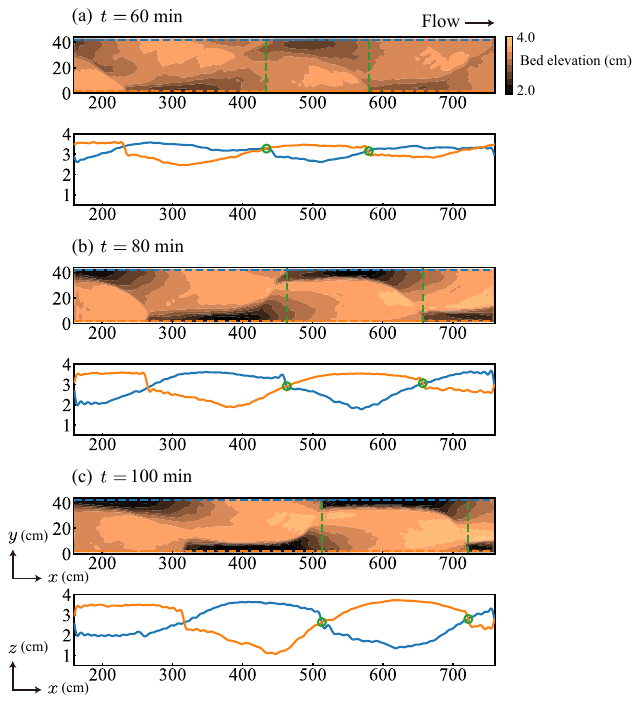}
 \caption{
 Temporal changes in bed elevation in plan view (top) and in longitudinal profiles (bottom) for Case 10, Replicate 1: (a) $t$ = 60 min, (b) $t$ = 80 min, and (c) $t$ = 100 min.
 The blue and orange dashed lines in the plan view correspond to the blue and orange solid lines in the longitudinal profiles, respectively. 
 The green circles in the longitudinal profiles indicate the intersection points of the left- and right-bank bed-elevation profiles and correspond to the green dashed lines in the plan view. 
 These intersections define the bar half-wavelength. 
 Within this half-wavelength, bar height is defined as the difference between the maximum and minimum bed elevations.
 }
 \label{zerocross}
\end{figure}

In this study, the height of alternate bars, $H_{\mathrm{MB}}$, was defined as the difference between the maximum and minimum bed elevations within a half-wavelength of alternate bar \cite{Ikeda1984, Redolfi2020}. 
The half-wavelength was chosen so that individual bars could be tracked for as long as possible within the available flume length. 
Because estimating bar height only requires that both the scour and depositional areas be included in the analysis domain, this choice is appropriate.
Figure \ref{zerocross} shows the definition of the half-wavelength based on the intersections of the longitudinal bed profiles extracted near the left and right sidewalls, where the characteristic alternate bar pattern is most clearly expressed.
The longitudinal section that most clearly expresses the bar waveform varied among runs.
We therefore computed the longitudinally averaged bed topography at the final time, and then selected, as representative longitudinal sections, the sections passing through the deepest points of the left- and right-bank scour zones, respectively.
In addition, dune-scale bedforms with wavelengths of approximately 5--10\ cm were present on the bars.
To extract the bar wavelength more accurately, we applied a filter to the longitudinal profiles that set the amplitudes of waves with wavelengths shorter than 10\ cm to zero \cite{Crosato2012}.
The analysis focused on a 6\ m reach between 3.6\ m and 9.6\ m downstream from the upstream end of the mobile-bed section. 
Among the bars that developed in this reach, we selected the downstream bar, where bed degradation due to a lack of sediment supply does not occur, and tracked its evolution in time.

\subsection{Estimation of Hydraulic Quantities over the Bars}
\label{method_cal}
The hydraulic quantities over the bars, required for interpreting the experimental results, were estimated by performing fixed-bed flow calculations in which the bed topography measured using ST was prescribed as the bed boundary condition. 
We used Nays2D in the iRIC package \cite{iRIC2016, iRIC2019}, which is a depth-averaged two-dimensional flow model based on the shallow-water equations.
Under supercritical flow conditions, the upstream discharge is prescribed as the boundary condition, whereas under subcritical flow conditions, the uniform flow depth at the downstream end is prescribed.
The roughness coefficient was set to 0.014\ m$^{-1/3}\cdot$s, computed from the mean grain size using the Manning-Strickler formula. 
The advection terms were discretized using the CIP (Cubic Interpolated Pseudo Particle) scheme. 
The computational grid consisted of square cells with a 1\ cm mesh size, identical to the spatial resolution of the ST measurements.
The time step was set to 0.005\ s.
The simulations were run for 60 s, by which time the hydraulic variables had become nearly steady.
%========================================================2============================================================

%========================================================3============================================================
\clearpage
\section{Experimental Results}
\label{result_exp}
This section presents the temporal evolution of bar height measured in the flume experiments and the corresponding growth rates inferred by nonlinear regression of the Landau equation.
We then examine how the inferred growth rate varies with discharge.

The overall process of bar initiation and development is outlined below. 
When flow was supplied over an initially flat bed, alternate bars formed under most hydraulic conditions.
Alternate bars then migrated downstream while largely preserving their planform geometry (Figure \ref{zerocross}). 
Under several conditions, multiple bar modes initially emerged, roughly ranging from 2 to 8 depending on the condition and stage of development; subsequently, progressive mode reduction ultimately led to a single alternate-bar pattern.
The imposed discharges were sufficient to submerge the bars under most conditions; however, at $Q$ = 1.7\ L/s, patches of the bed surface on the order of a few centimeters were exposed in the upstream reach during the late stage of some runs.

In Case 16, a bed configuration that could be regarded as alternate bars formed, but its morphology was influenced by the block placed at the upstream end. 
This case has the smallest difference between $\tau_{*0}$ and $\tau_{*\mathrm{c}}$ among the 16 conditions (Table \ref{condition}).
As a result, sediment transport was weak, and bar height did not reach its equilibrium value.
This case therefore is excluded from the analysis in the present study.
We also excluded replicates that could not be tracked for extended periods (e.g., due to disruption of the alternate-bar pattern) and replicates that showed substantially lower reproducibility than the others.

\subsection{Temporal Evolution of the Alternate-Bar Height}
\begin{figure}[t]
 \centering
   \includegraphics[width=1.0\textwidth]{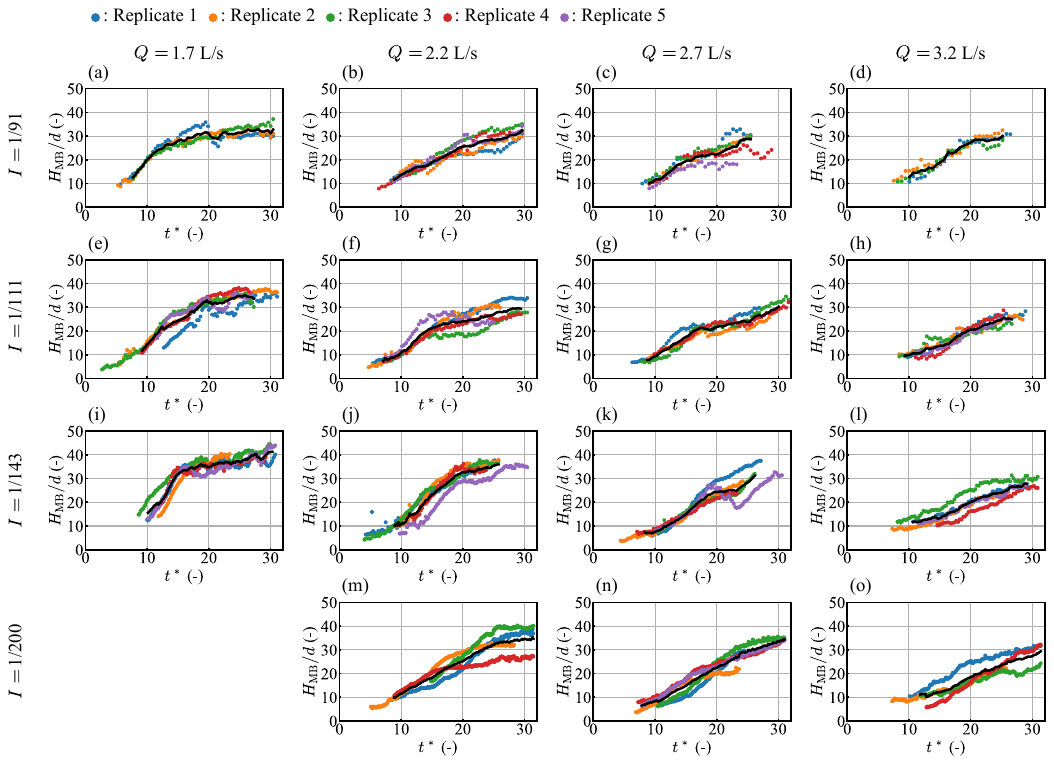}
 \caption{
 Temporal evolution of the alternate-bar height. 
 Rows correspond to channel slope ($I$ = 1/91, 1/111, 1/143, and 1/200, from top to bottom), and columns correspond to discharge ($Q$ = 1.7, 2.2, 2.7, and 3.2\ L/s, from left to right).
 The black solid lines indicate the mean values across replicates.
 The growth speed of bar height is low during the early stage of the run, reaches a maximum in the middle stage, and then decreases toward the end as the system approaches equilibrium.
 Both the growth speed and the equilibrium value tend to be larger under lower-discharge conditions.
 }
 \label{height}
\end{figure}

Figure \ref{height} shows the temporal evolution of the alternate-bar height.
Bar height is often nondimensionalized using flow depth; but we scale it by the bed-material grain size $d$ because our primary objective is to compare bar height across discharges \cite{Redolfi2020}.
The figure shows that the temporal evolution of bar height follows an approximately sigmoidal curve across all conditions: the time derivative of bar height (growth speed) is largest during the middle of the run (15 $<$ $t^*$ $<$ 20), then decreases toward the end (20 $<$ $t^*$) as the bars approach an equilibrium state.
Expressed in terms of the normalized bar height $H_{\mathrm{MB}}/d$, the period of rapid bar growth corresponds to approximately 15 $<$ $H_{\mathrm{MB}}/d$ $<$ 30.
It is also qualitatively apparent that, for most channel slopes, lower discharge generally tends to produce faster growth on a bedload-based timescale and a larger equilibrium bar height.
However, this tendency is weak under the $I$ = 1/91 condition, whereas it is particularly evident under the $I$ = 1/143 condition.
As a side note, bar wavelength had already become nearly constant during the period when bar-height growth was most pronounced \cite{Fujita1985} under all conditions.
The wavelength ultimately converged to a value of approximately eight times the channel width, and its dependence on discharge was much weaker than that of bar height.
These are because the bar wavelength is primarily controlled by channel width. 

\begin{figure}[t]
 \centering
   \includegraphics[width=1.0\textwidth]{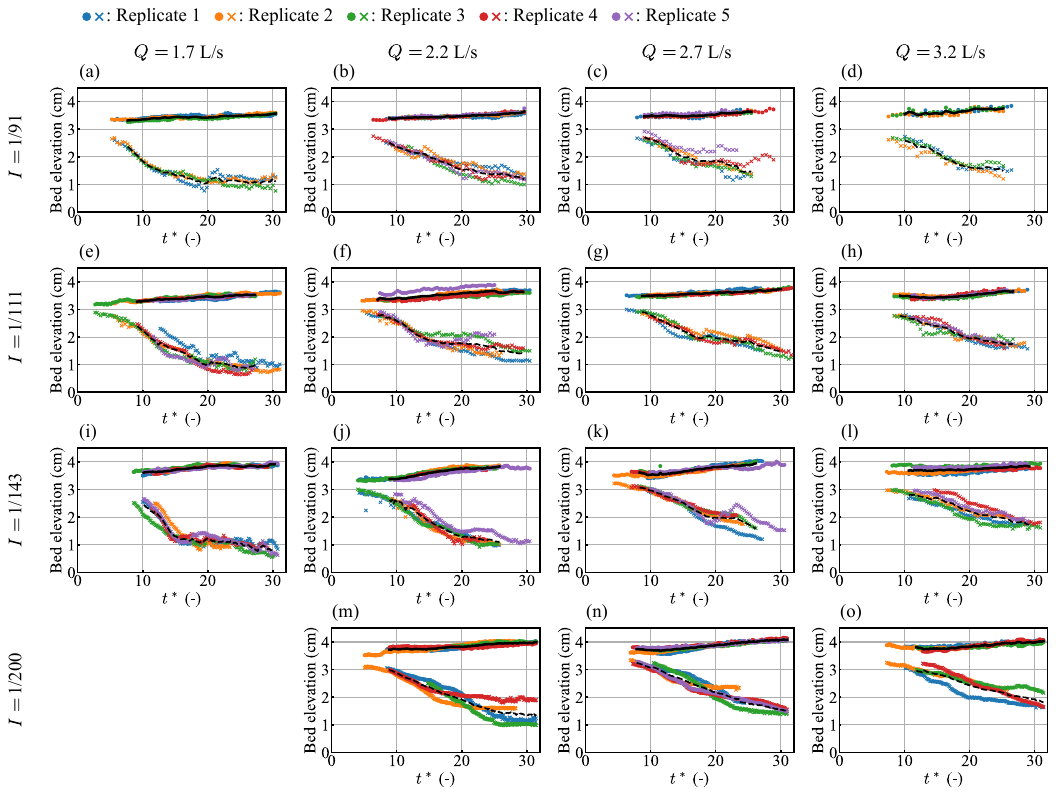}
 \caption{
 Temporal changes in the deposition and scour locations.
 Rows correspond to channel slope ($I$ = 1/91, 1/111, 1/143, and 1/200, from top to bottom), and columns correspond to discharge ($Q$ = 1.7, 2.2, 2.7, and 3.2\ L/s, from left to right).
 Circles and crosses indicate the maximum and minimum bed elevations, respectively.
 The black solid line shows the replicate-mean maximum bed elevation, and the black dashed line shows the replicate-mean minimum bed elevation.
 The growth of bar height is governed primarily by scour rather than deposition.
 }
 \label{dpst_scr}
\end{figure}

Here, we examine which of bedload deposition and scour primarily controls the growth of bar height.
Figure \ref{dpst_scr} shows the temporal changes in the maximum and minimum bed elevations. 
The maximum bed elevation shows only a slight increasing trend, whereas the minimum bed elevation continues to decrease with a slope similar to that of the bar height and eventually converges to an equilibrium value.
This result suggests that the increase in bar height observed in our experiments is primarily attributable to progressive bed scour.
This is consistent with the findings of \citeA{Fujita1985}.
According to their study, the scoured sediment is consumed in the transverse protrusion of the bar front in response to the divergence of meandering flow, causing the scour zone to become narrower.
The contour map in Figure \ref{zerocross} shows exactly this behavior, indicating that the phenomenon observed in the experiments of \citeA{Fujita1985} was also present in our experiments.
In Section \ref{mechanism}, under this assumption, we discuss the physical mechanisms by which the development of bar height depends on discharge.

%=======================================================Figs==========================================================

%=======================================================Figs==========================================================

\subsection{Temporal Changes in Bar Celerity}
Bar mobility exerts a primary control on how readily bar height grows.
During large floods, bars migrate downstream; however, depending on flood magnitude, the depositional areas may become emergent, preventing bar migration.
Under such conditions, the flow and associated sediment transport become locally concentrated near the banks, accelerating bed degradation and promoting the development of bar height.
This section examines whether this phenomenon occurs in our experimental system.

\begin{figure}[t]
 \centering
   \includegraphics[width=1.0\textwidth]{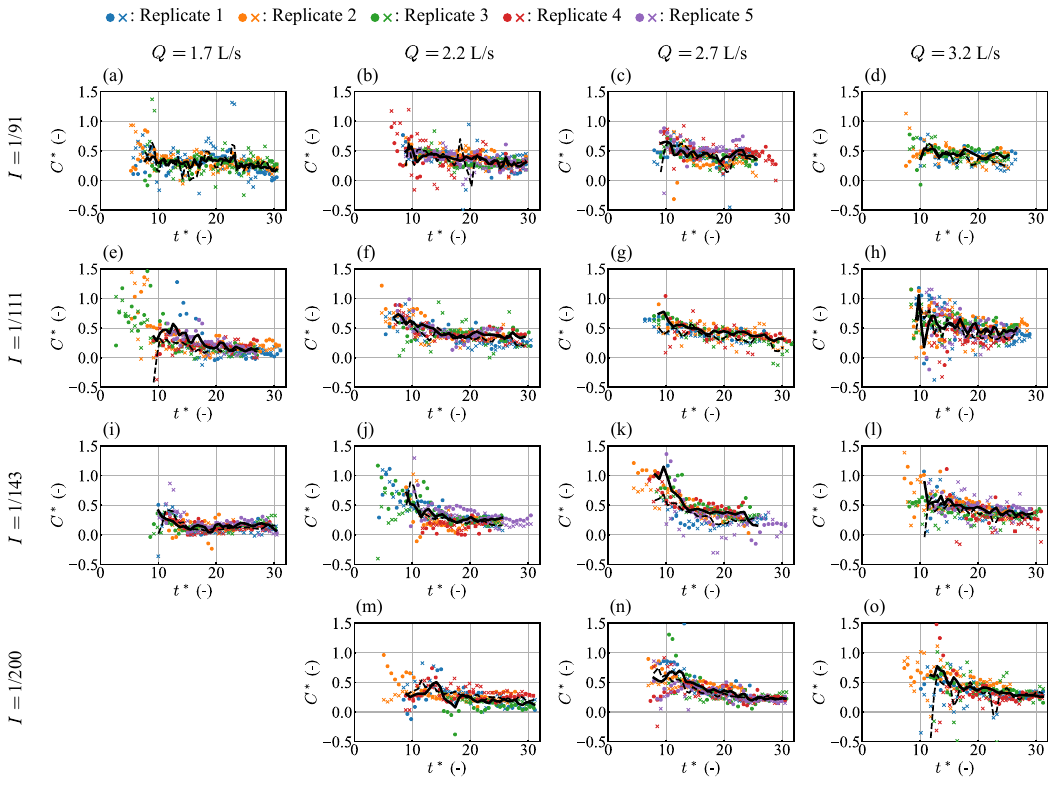}
 \caption{
 Temporal changes in the bar celerity. 
 Rows correspond to channel slope ($I$ = 1/91, 1/111, 1/143, and 1/200, from top to bottom), and columns correspond to discharge ($Q$ = 1.7, 2.2, 2.7, and 3.2\ L/s, from left to right).
 Circles and crosses indicate the upstream and downstream ends of the target bar, respectively.
 The black solid line shows the replicate-mean position of the upstream end, and the black dashed line shows the replicate-mean position of the downstream end.
 For convenience in estimating bar celerity, the values were subsampled from the full dataset.
 The dimensionless bar celerity $C^*$ increases with discharge.
 }
 \label{speed}
\end{figure}

\begin{figure}[t]
 \centering
   \includegraphics[width=1.0\textwidth]{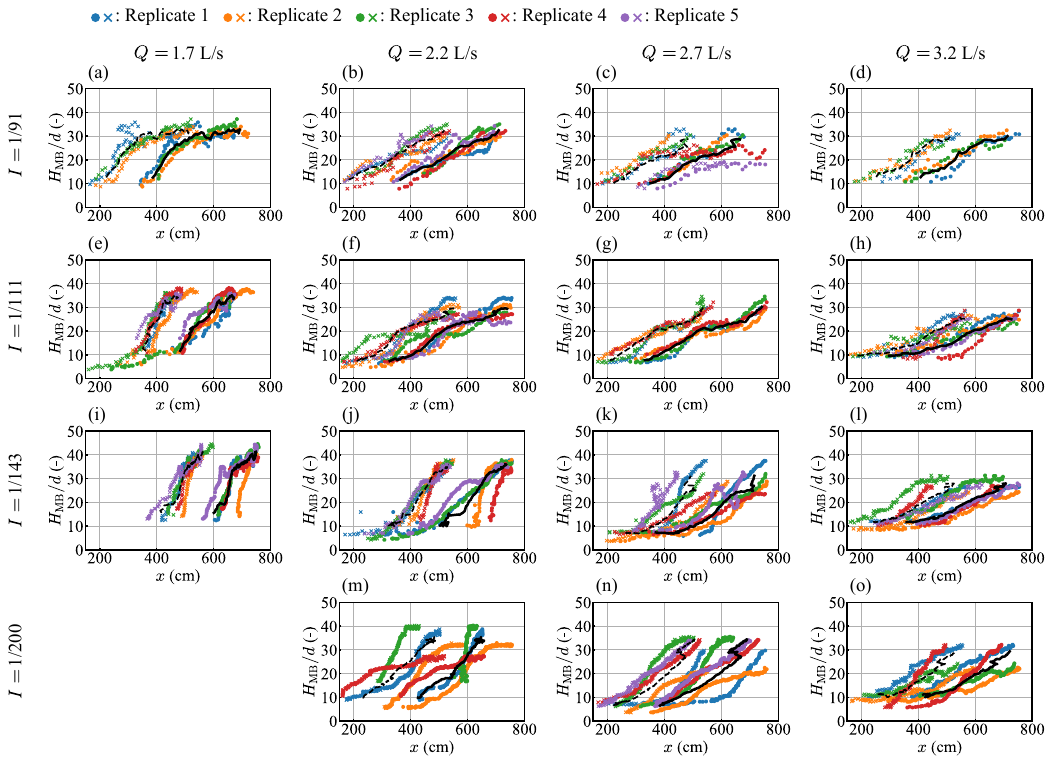}
 \caption{
 Relationship between bar position and bar height.
 Rows correspond to channel slope ($I$ = 1/91, 1/111, 1/143, and 1/200, from top to bottom), and columns correspond to discharge ($Q$ = 1.7, 2.2, 2.7, and 3.2\ L/s, from left to right).
 Circles and crosses indicate the upstream and downstream ends of the target bar, respectively.
 The black solid line shows the replicate-mean position of the upstream end, and the black dashed line shows the replicate-mean position of the downstream end.
 The bar height is the same as that shown in Figure \ref{height}.
 Under low-discharge conditions, bars grow in height with almost no downstream migration, whereas under high-discharge conditions, they grow while migrating downstream.
 }
 \label{height_vs_x}
\end{figure}

Figure \ref{speed} shows the temporal evolution of bar celerity. 
The dimensionless celerity $C^*$ is defined as the time derivative of the normalized bar position $x/B$. 
This position is the streamwise location of the intersection points between the longitudinal bed-elevation profiles extracted near the left and right sidewalls (green circles in Figure \ref{zerocross}).
$C^*$ is given by the following expression: 
%%%%%%%%%%%%%%%%%%%%%%%%%%%%%%%%%%%
\begin{equation}
C^* = \frac{d(x/B)}{dt^*}.
\end{equation}
%%%%%%%%%%%%%%%%%%%%%%%%%%%%%%%%%%%
The figure shows that $C^*$ tends to decrease with time under all conditions. 
This tendency is likely related to the temporal evolution of bars, in which the wavelength lengthens \cite<e.g.,>{Crosato2012} and the bar height increases over time.
It is also found that the mean value of $C^*$ is smaller for lower discharges. 
This tendency is particularly evident for the conditions with $I$ = 1/111 and $I$ = 1/143, in which the dependence of the bar growth rate on discharge was most pronounced. 
For example, at $I$ = 1/143, the value of $C^*$ for $Q$ = 1.7\ L/s is approximately half of that for $Q$ = 3.2\ L/s. 
Furthermore, Figure \ref{height_vs_x} shows the relationship between bar position $x$ and bar height, using the same bar height as in Figure \ref{height}.
During the stage of bar-height growth, under lower discharges, bars grow in height with almost no downstream migration, whereas under higher discharges they grow in height while migrating downstream.
Taken together, these results suggest that lower discharges tend to produce flow conditions in which bar migration is more strongly suppressed. 

%=======================================================Figs==========================================================

%=======================================================Figs==========================================================

\clearpage
\subsection{Quantification of the Relationship Between Growth Rate of Bar Height and Water Discharge}

\subsubsection{Application of the Landau Equation}
\label{appl_LE}
According to weakly nonlinear stability theory for bars, the temporal evolution of the amplitude $\alpha_{11}$ of the dominant bar mode is expected to follow the Landau equation \cite <e.g.,>{Colombini1987, Tubino1991, Schielen1993, Hall2004, Bertagni2018, Ali2021}:
%%%%%%%%%%%%%%%%%%%%%%%%%%%%%%%%%%%
\begin{equation}
\frac{d\alpha_{11}}{dt} = \Omega_{11} \alpha_{11}
\left\{
1 - \left(\frac{\alpha_{11}} { {\alpha_{11\mathrm{e}}} }\right)^2
\right\},
\label{Landau}
\end{equation}
%%%%%%%%%%%%%%%%%%%%%%%%%%%%%%%%%%%
where $\Omega_{11}$ is the linear growth rate, and $\alpha_{11\mathrm{e}}$ is the equilibrium amplitude. 
Equation (\ref{solution}) represents a generic functional form describing the growth and saturation of a physical quantity. 

\begin{figure}[t]
 \centering
   \includegraphics[width=1.0\textwidth]{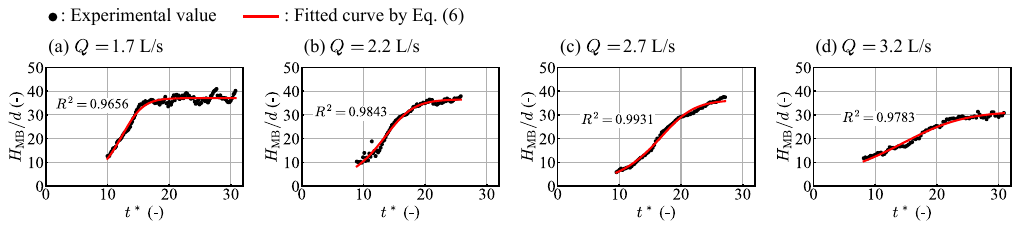}
 \caption{
 Examples of the Landau-equation fit for $I=1/143$ at each discharge.
 }
 \label{Landau_fit}
\end{figure}

Because bars are a superposition of multiple Fourier components, there is no theoretical guarantee that bar height follows a Landau equation.
However, the overall shape of Figure \ref{height} suggests that bar height in our experimental system exhibits nonlinear behavior that can be described by a Landau equation.
Based on this assumption, we make an approximation by replacing $\alpha_{11}$ in Equation (\ref{solution}) with $H_{\mathrm{MB}}$, so that
%%%%%%%%%%%%%%%%%%%%%%%%%%%%%%%%%%%
\begin{equation}
\frac{dH_{\mathrm{MB}}}{dt} = \Omega_{H_{\mathrm{MB}}} H_{\mathrm{MB}}
\left\{
1 - \left(\frac{H_{\mathrm{MB}}} { H_{\mathrm{MBe}} }\right)^2
\right\},
\label{Landau_height}
\end{equation}
%%%%%%%%%%%%%%%%%%%%%%%%%%%%%%%%%%%
where $ \Omega_{H_{\mathrm{MB}}} $ is the growth rate of bar height, and $H_{\mathrm{MBe}}$ is the equilibrium bar height.
The general solution of Equation (\ref{Landau_height}) can be written as
%%%%%%%%%%%%%%%%%%%%%%%%%%%%%%%%%%%
\begin{equation}
H_{\mathrm{MB}}(t) = \frac{H_{\mathrm{MBe}}}
{\sqrt{1+\dfrac{{H_{\mathrm{MBe}}}^2-{H_{\mathrm{MB0}}}^2}{{H_{\mathrm{MB0}}}^2}e^{-2\Omega_{H_{\mathrm{MB}}} t}}},
\label{solution}
\end{equation}
%%%%%%%%%%%%%%%%%%%%%%%%%%%%%%%%%%%
where $H_{\mathrm{MB0}}$ denotes the value of $H_{\mathrm{MB}}$ at $t$ = 0.
Figure \ref{Landau_fit} shows an example in which Equation (\ref{solution}) was fitted to the measured bar-height time series using a nonlinear least-squares method.
The fitting accuracy of the Landau equation is discussed in detail in Appendix \ref{LEF}.
This procedure uniquely determines $\Omega_{H_{\mathrm{MB}}}$, $H_{\mathrm{MBe}}$ and $H_{\mathrm{MB0}}$. 
In this study, $\Omega_{H_{\mathrm{MB}}}$ was scaled using Equation (\ref{t*}).
$H_{\mathrm{MBe}}$ was scaled by flow depth for comparison with previous studies (Appendix \ref{EBH}).

\subsubsection{Sensitivity of Bar Growth Rate on Discharge}
\begin{figure}[t]
 \centering
   \includegraphics[width=1.0\textwidth]{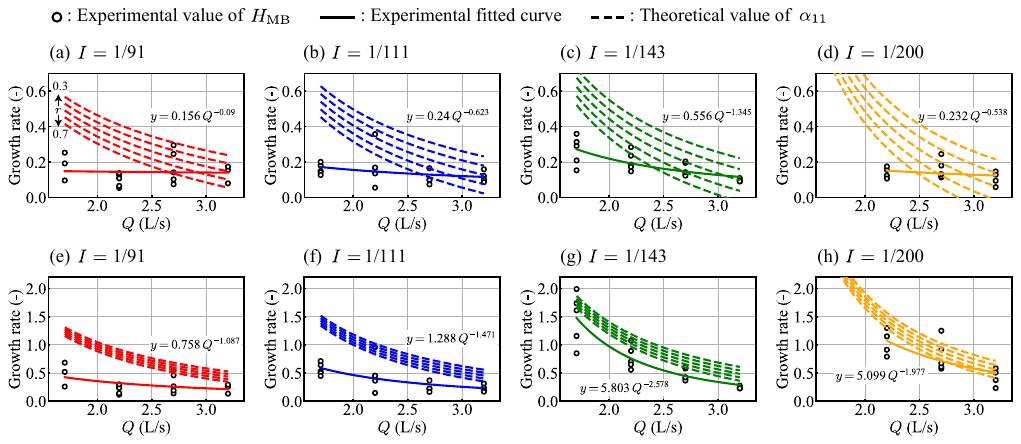}
 \caption{
 Relationship between the bar growth rate and water discharge.
 Panels (a)--(d) show the results obtained using the sediment transport formula of \citeA{mpm} at each channel slope, whereas panels (e)--(h) show those obtained using the formula of \citeA{Parker1978} at each channel slope.
 The theoretical values in this range increase monotonically with decreasing discharge, so we fitted the relationship between the experimental values $y$ and discharge using a simple monotonic power-law function $y$ = $b$ $\cdot$ $Q^a$, and plotted the regression as a solid line.
 The sensitivity to discharge can be assessed from the value of $a$. 
 The measured growth rate of bar height is less sensitive to discharge than the linear growth rate of the amplitude of a single alternate-bar mode predicted by linear stability theory.
 }
 \label{growthrate}
\end{figure}

Figure \ref{growthrate} shows the relationship between the dimensionless growth rate of bar height $\Omega_{H_{\mathrm{MB}}}$ and water discharge.
As a reference for the experimental values obtained in this study, the theoretical values $\Omega_{11}$ predicted by linear stability analysis are shown by the dashed lines. 
This value is the maximum growth rate of the transverse mode $m$ = 1 of alternate bars, and the analysis followed the approach of \citeA{Colombini1987}.

Because the analytical results were expected to vary depending on the sediment transport formula, we employed two different formulas.
One is the \citeA{mpm} formula (\ref{MPM}), which is widely used internationally, and the other is the \citeA{Parker1978} formula shown below, which has been used in previous theoretical studies \cite{Redolfi2021, Carlin2021, Redolfi2023} and has a different functional form from the \citeA{mpm} formula.
%%%%%%%%%%%%%%%%%%%%%%%%%%%%%%%%%%%
\begin{equation}
q_{\mathrm{s}0} = 11.2\frac{(\tau_{*0}-\tau_{*\mathrm{c}})^{9/2}}{{\tau_{*0}}^3}\sqrt{sgd^3},
\end{equation}
%%%%%%%%%%%%%%%%%%%%%%%%%%%%%%%%%%%
Note that both formulas give $\tau_{*\mathrm{c}}$ = 0.034.
The time in the analysis was nondimensionalized using Equation (\ref{t*}), in the same manner as in the experiments.
The empirical constant $r$ in the relation that determines the direction of bedload transport (see, e.g., Equation (11) of \citeA{Colombini1987}) is generally taken to be between 0.3 and 0.7.
However, because bedload transport in the $y$ direction is generally difficult to measure, we plot the full range of values from 0.3 to 0.7 here.
We also tested both the \citeA{Engelund1967} formula, which is frequently used in stability analyses, and the Manning formula for the bed-friction coefficient $C_\mathrm{f}$.
Because no clear difference was found between the two, only the results obtained using the \citeA{Engelund1967} formula are shown.
The \citeA{Engelund1967} formula is given as follows.
%%%%%%%%%%%%%%%%%%%%%%%%%%%%%%%%%%%
\begin{equation}
C_{\mathrm{f}}^{-\frac{1}{2}} = 6+2.5\ln
\left(
\frac{h_0}{2.5d}
\right)
\end{equation}
%%%%%%%%%%%%%%%%%%%%%%%%%%%%%%%%%%%
Note that $Q_{\mathrm{c}}$ in Table \ref{condition} was calculated using the \citeA{mpm} formula, the \citeA{Engelund1967} formula, and $r$ = 0.5.

The theoretical predictions show that, over the range of discharges considered, the dimensionless linear growth rate increases monotonically with decreasing discharge on a sediment-transport-based timescale.
Turning to the experimental results, the values of $a$ are negative under all conditions, indicating the same tendency as in the theoretical predictions; namely, the dimensionless growth rate increases with decreasing discharge.
However, the experimental response to discharge is weaker than that predicted by theory, except for the $I$ = 1/143 and $I$ = 1/200 cases when the Parker formula is used.
In particular, the response is nearly constant for $I$ = 1/91.
The discrepancy between the experimental and theoretical values tends to become larger under lower-discharge conditions.
As shown in Appendix \ref{STF}, however, the reproducibility of sediment transport in our experiments is better represented by the \citeA{mpm} formula.
Therefore, the results shown in panels (a)--(d) are considered to be more reliable.
In addition, varying $r$ mainly changes the magnitude of the growth rate, without substantially affecting its dependence on discharge.
Based on the comparison between the analytical and experimental growth rates, the \citeA{mpm} formula with $r$ = 0.5 and the \citeA{Parker1978} formula with $r$ = 0.7 appear to provide the closest agreement with the experimental results.

These results indicate that the measured growth rate of alternate-bar height is less sensitive to discharge than predicted by linear stability analysis.
In other words, the linear growth rate predicted for a single Fourier component does not adequately approximate the discharge sensitivity of the growth rate of bar height.
The possible reasons for this discrepancy and implications for extending the theoretical framework are discussed in Section \ref{factors}.

%========================================================3============================================================

%========================================================4============================================================
\clearpage
\section{Hydraulic Calculation Results}
\label{result_cal}
In the preceding section, we quantified the response of the dimensionless growth rate of alternate-bar height to water discharge by ST measurement.
The results suggest a general tendency for this growth rate to increase with decreasing discharge on a dimensionless timescale based on bedload transport.
Previous experimental studies have reported related phenomena.
For example, \citeA{Redolfi2020} showed that the equilibrium bar height increases with decreasing discharge under fully wet conditions, whereas \citeA{Visconti2010} observed progressive bank erosion in response to decreasing discharge.
%out: Related phenomena---such as an equilibrium bar height that peaks at an intermediate discharge \cite{Redolfi2020} and progressive bank erosion when the imposed discharge is reduced \cite{Visconti2010}---have been reported in previous experimental studies.
In contrast, much of this work has been limited to reporting qualitative experimental results, and the physical factors controlling these phenomena have not been discussed in sufficient detail.
One hypothesis that may explain the monotonic increase in dimensionless growth rate with decreasing discharge is as follows. 
Under lower-discharge conditions, two processes occur during bar development.
First, flow deflection becomes more pronounced under the influence of gravity.
Second, sediment transport is suppressed immediately downstream of the depositional area, reducing bar mobility.
These processes concentrate flow and sediment transport in the scour zones, thereby enhancing local scour and promoting the growth of bar height
In this section, to examine this hypothesis, we focus on quantities---flow deflection and the extent of the non-sediment-transport zones---and investigate their relationships with the growth of bar height.

\subsection{Flow Deflection and Non-Sediment-Transport Zones} 
Here we investigate how the structure of the flow field influences the growth of bar height
Specifically, we examine how the flow-deflection index $v/u$, defined as the ratio of transverse to streamwise velocity, and the Shields number $\tau_*$, which characterizes sediment-transport intensity, vary with discharge.

\begin{figure}[t]
 \centering
   \includegraphics[width=1.0\textwidth]{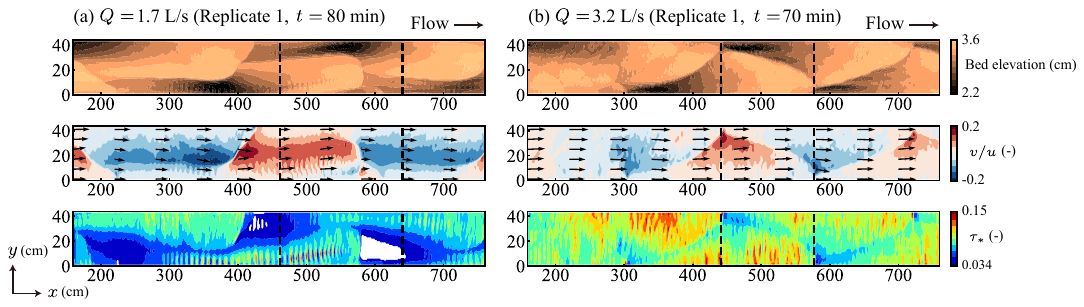}
 \caption{
 Contour maps at $H_\mathrm{MB}/d$ $\approx$ 20 for $I$ = 1/143.
 The first, second, and third rows show bed elevation, $v/u$, and $\tau_*$, respectively.
 (a) Case 12, Replicate 1, at 80 min; (b) Case 9, Replicate 1, at 70 min. 
 The target bar is outlined by the black dashed line.
 In the $v/u$ pales, positive $v$ is directed toward the left bank, and the arrows indicate streamlines.
 In the $\tau_*$ panels, the white shading indicates zones where $\tau_*$ $<$ $\tau_{*\mathrm{c}}$ = 0.034.
 Under low-discharge conditions, flow deflection is pronounced, non-sediment-transport zones develop, and sediment transport is localized within the scour zone.
 Under high-discharge conditions, the flow is strongly aligned with the channel, and sediment transport occurs over the entire bar surface.
}
 \label{contour}
\end{figure}

\begin{figure}[t]
 \centering
   \includegraphics[width=1.0\textwidth]{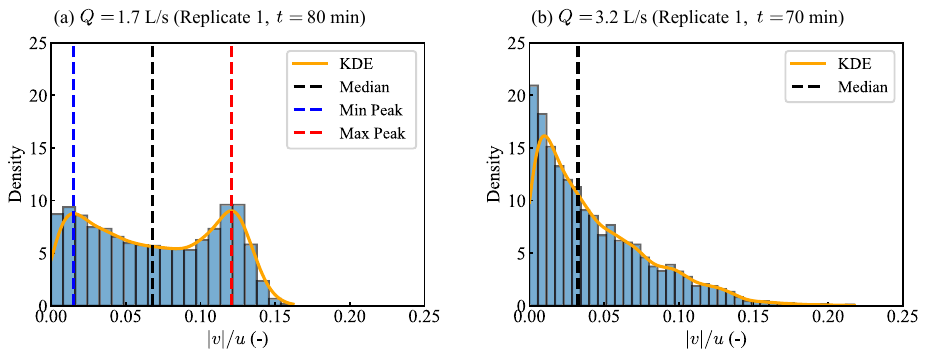}
 \caption{
 Frequency distribution of $|v|/u$. 
 (a) Case 12, Replicate 1, at 80 min; (b) Case 9, Replicate 1, at 70 min. 
 The orange line shows a kernel density estimate (KDE) of the distribution.
 The distribution is multimodal (here, bimodal) under low-discharge conditions, whereas it is unimodal under high-discharge conditions.
 }
 \label{hist}
\end{figure}

Figure \ref{contour} shows the spatial distributions of bed elevation, $v/u$, and $\tau_*$ when $H_\mathrm{MB}/d$ $\approx$ 20, i.e., during the stage of rapid bar growth (Figure \ref{height}).
In the panels for $\tau_*$, zone where $\tau_*<\tau_{*\mathrm{c}}$ are shown in white to visualize the non-sediment-transport zones.
As representative examples, results are presented for the conditions with $I$ = 1/143, $Q$ = 3.2\ L/s, and $Q$ = 1.7\ L/s. 
Figure \ref{hist} shows the frequency distribution of $|v|/u$ of these conditions.
The $v/u$ panels in Figure \ref{contour} show that, compared with $Q$ = 3.2\ L/s, the $Q$ = 1.7\ L/s condition exhibits larger $v/u$ values near the bar front, indicating a more strongly deflected flow. 
In particular, the transverse component reaches values up to about 20\ \% of the streamwise component.
Figure \ref{hist} shows that the frequency distribution is unimodal for $Q$ = 3.2\ L/s, whereas it is bimodal for $Q$ = 1.7\ L/s.
The $\tau_*$ panels in Figure \ref{contour} show that, whereas sediment transport occurs across almost the entire channel for $Q$ = 3.2\ L/s, transport ceases immediately downstream of the depositional areas for $Q$ = 1.7\ L/s. 

\begin{figure}[t]
 \centering
   \includegraphics[width=0.7\textwidth]{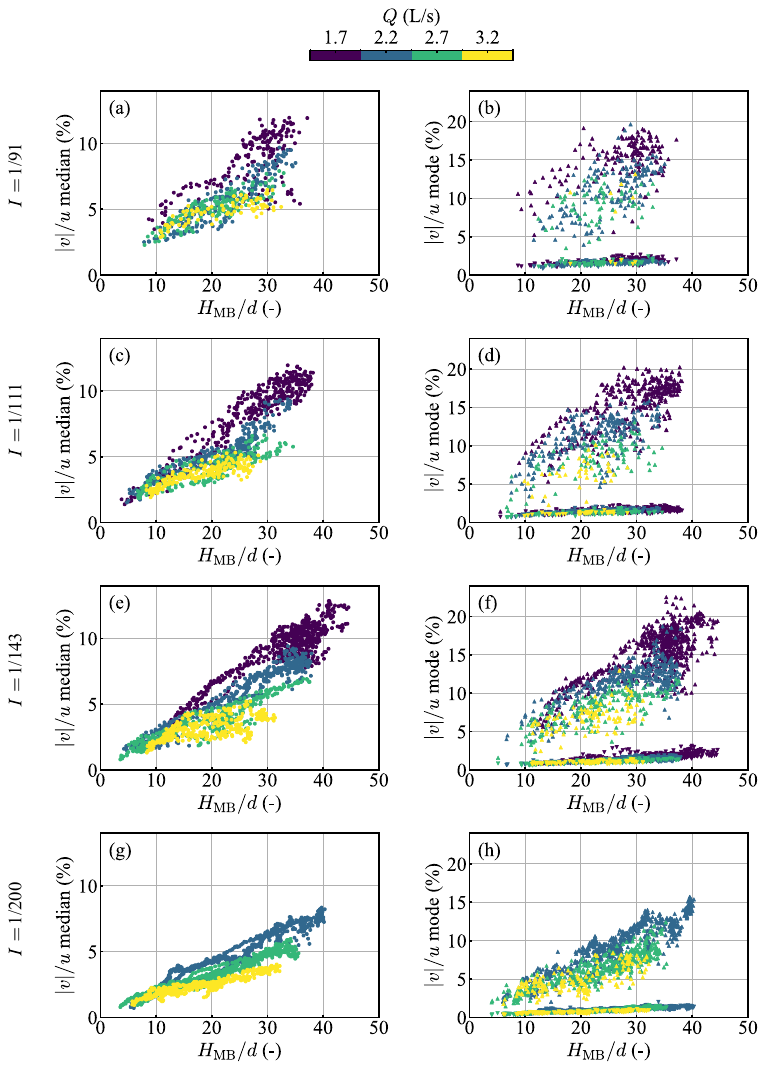}
 \caption{
 Representative values of $|v|/u$, an index of flow deflection on the target bar.
 The left panels (a), (c), (e), and (g) show the median (circles), and the right panels (b), (d), (f), and (h) show the mode (triangles).
 Rows correspond to channel slope ($I$ = 1/91, 1/111, 1/143, and 1/200, from top to bottom).
 Note that the vertical-axis range differs between the median and the mode.
 Upward triangles indicate the largest peak of a multimodal distribution, and downward triangles indicate the smallest peak.
 Flow deflection becomes stronger as discharge decreases and bar height increases. 
 }
 \label{flow}
\end{figure}

\begin{figure}[t]
 \centering
   \includegraphics[width=0.7\textwidth]{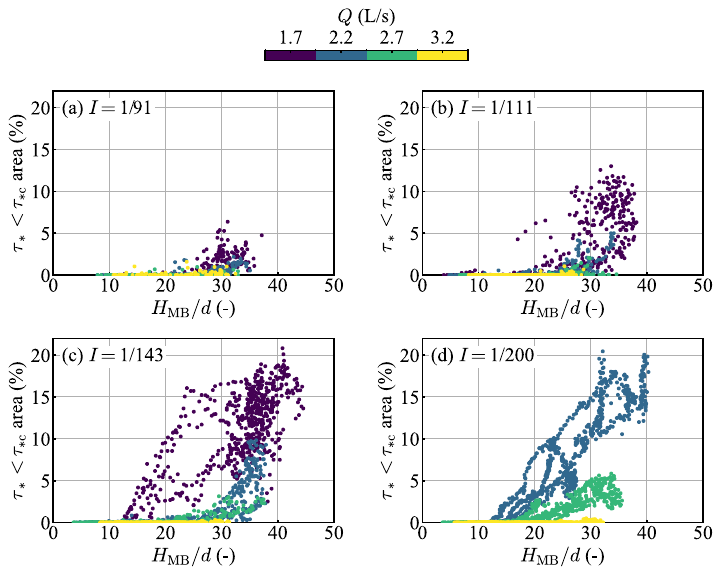}
 \caption{
 Fractional area of the non-sediment-transport zones on the target bar at each channel slope.
 These zones expand as discharge decreases and bar height increases. 
 }
 \label{sediment}
\end{figure}

To further generalize the results with respect to discharge, Figure \ref{flow} shows representative spatial values of $|v|/u$, and Figure \ref{sediment} shows the areal fraction of the zone where $\tau_*$ $<$ $\tau_{*\mathrm{c}}$, both plotted as functions of the height-to-diameter ratio. 
Using the height-to-diameter ratio as the horizontal axis allows comparison among discharges at comparable stages of bar development, that is, when the bars have similar heights.
For $|v|/u$, the left column of Figure \ref{flow} presents the median of the spatial samples.
However, at lower discharge, the frequency distribution of $|v|/u$ tends to be multimodal rather than unimodal (Figure \ref{hist}), so the median alone is not always sufficient to characterize the distribution.
Therefore, we also evaluated modal values from the kernel density function of the distribution.
Specifically, when multiple peaks were detected, the maximum and minimum peak values were adopted as the modal values and are shown in the right column of Figure \ref{flow}.

Figure \ref{flow} shows that, for all channel slopes, the median and mode of $|v|/u$ increase as discharge decreases at a given $H_{\mathrm{MB}}/d$.
For example, for $I$ = 1/143 at $H_{\mathrm{MB}}/d$ = 30, the median is up to about two times larger and the mode up to about 1.5 times larger at $Q$ = 1.7 L/s than at $Q$ = 3.2 L/s.
Figure \ref{sediment} shows that, for all channel slopes, the area of the non-transport zones also tends to increase as discharge decreases at a given $H_{\mathrm{MB}}/d$. 
For $I$ = 1/143 at $H_{\mathrm{MB}}/d$ = 30, the non-transport zones at $Q$ = 1.7\ L/s is more than 15 times that at $Q$ = 3.2\ L/s. 
Overall, these results indicate that flow deflection and expansion of the non-transport zones become more pronounced as discharge decreases and bar height increases (i.e., as the height-to-depth ratio increases). 

%=======================================================Figs==========================================================

%=======================================================Figs==========================================================

\subsection{Effects of Bar Shape on Flow Deflection}
It has been reported that the shape of alternate bars varies with discharge magnitude \cite{Redolfi2020}.
In our experiments (see the bar morphology in Figure \ref{contour}), a similar tendency was observed:
for $Q$ = 1.7\ L/s it tended to bulge in the cross-channel direction, whereas for $Q$ = 3.2\ L/s the bar front was nearly straight, thereby supporting the findings of \citeA{Redolfi2020}. 
Is the enhanced flow deflection over bars controlled by an increase in the bar height-to-depth ratio, or is it simply a consequence of differences in bar shape?

\begin{figure}[t]
 \centering
   \includegraphics[width=0.75\textwidth]{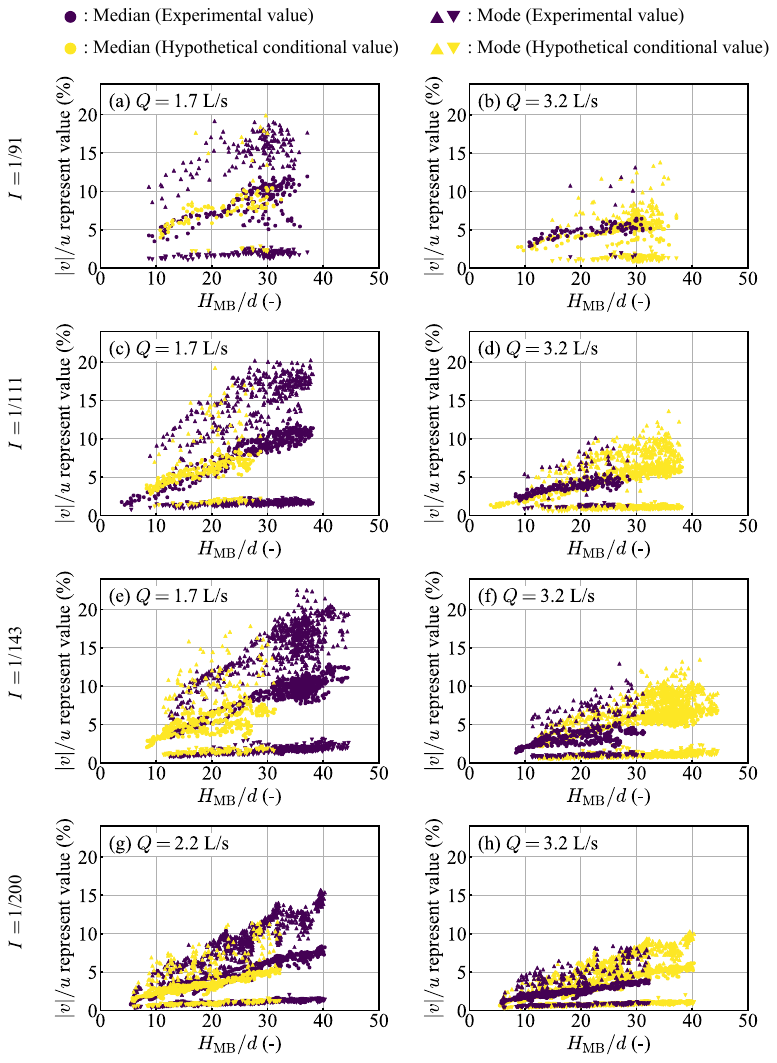}
 \caption{
 Representative values of $|v|/u$, a measure of flow deflection considering bar shape alone.
 The left panels (a), (c), (e), and (g) show the smaller discharge, and the right panels (b), (d), (f), and (h) show the larger discharge. 
 Rows correspond to channel slope ($I$ = 1/91, 1/111, 1/143, and 1/200, from top to bottom).
 Upward triangles indicate the largest peak of a multimodal distribution, and downward triangles indicate the smallest peak.
 Flow deflection is more strongly influenced by the height-to-depth ratio than by bar shape.
 }
 \label{flow_shape}
\end{figure}

To address this question, we additionally performed fixed-bed simulations in which a hypothetical discharge of $Q$ = 1.7\ L/s was applied to the bed topographies obtained at each time step in the $Q$ = 3.2\ L/s experiments.
Likewise, we carried out fixed-bed simulations in which a hypothetical discharge of $Q$ = 3.2\ L/s was imposed on the bed topographies formed under the $Q$ = 1.7\ L/s experiments. 
The results are shown in Figure \ref{flow_shape}. 
In each panel, the conditions are arranged by channel slope and discharge, allowing comparisons in which bar shape is the only variable. 
Figure \ref{flow_shape} shows that for all combinations of channel slope and discharge, the differences in $|v|/u$ caused by variations in bar shape are smaller than those caused by differences in discharge.
This suggests that the height-to-depth ratio is more influential than bar shape for the flow deflection. 

%========================================================4============================================================

%========================================================5============================================================
\clearpage
\section{Discussion}

\subsection{Factors Contributing to the Discrepancy Between Experimental and Theoretical Bar Growth Rates}
\label{factors}
Section \ref{result_exp} showed that the experimentally inferred growth rate of alternate-bar height exhibits a weaker response to discharge than the theoretical prediction for the amplitude of a single alternate-bar mode (Figure \ref{growthrate}).
This discrepancy becomes particularly pronounced under low-discharge conditions.
In this section, to clarify the range of applicability of the theory and to improve its predictive accuracy, we discuss the causes of the discrepancy between experiments and theory from two perspectives:
(i) the validity range of the linearization adopted in linear stability analysis, and (ii) interactions among multiple wavenumber components in the experimental system.
We further describe how a theory that resolves these issues could contribute to predicting bed deformation in natural rivers.

\subsubsection{Limitations of the Linearized Solution}
\label{lim}
Linear stability analysis of bars linearizes the solutions of the coupled system consisting of the two-dimensional shallow-water equations, the fluid continuity equation, and the Exner equation.
Perturbations are then imposed on departures from spatially uniform flow, assuming that fluctuations in velocity, flow depth, and bed elevation are of small amplitude (i.e., one order of magnitude smaller than the base state).
When flow nonuniformity and bed relief become pronounced, this assumption may break down, and the predictive accuracy of the analysis is expected to deteriorate.

In Section \ref{result_cal}, we showed that, under low-discharge conditions, the height-to-depth ratio is relatively large, the flow over bars exhibits deflection, and sediment transport is spatially heterogeneous.
As an example of the flow response, at low discharge, the transverse velocity near the bar front reaches as much as 20\ \% of the streamwise velocity (Figure \ref{contour}).
Moreover, the distribution of $|v|/u$ is bimodal rather than unimodal, suggesting two distinct deflection regimes and pronounced spatial variability (Figure \ref{hist}).
Under low-discharge conditions, that is, when the width-to-depth ratio is large, these features indicate larger departures from the idealized uniform-flow assumption, which likely explains why the discrepancy between theoretical and experimental growth rates is most pronounced.
This constitutes a major limitation of stability analysis, because even when extended to weakly nonlinear analysis, its range of applicability remains restricted to conditions very close to the critical aspect ratio.

\subsubsection{Role of Interactions among Multiple Bar Modes in the Nonlinear Growth of Alternate Bars}
\label{multi-mode}
In stability analyses of alternate bars, the most amplified component of the $m$ = 1 mode---namely the longitudinal mode $n$ = 1---is often assumed to dominate.
This assumption neglects interactions with other wavenumber components.
However, studies applying two-dimensional Fourier transforms to alternate-bar morphology have pointed out that the amplitudes of other components $\alpha_{mn}$, such as ($m,\ n$) = (2, 0), can be of the same order as that of the ($m,\ n$) = (1, 1) component \cite{Hasegawa1982, Garcia1993, Redolfi2020}.
\citeA{Redolfi2020} also reported that the ($m,\ n$) = (2, 0) and ($m,\ n$) = (3, 1) components become more dominant under relatively low-discharge conditions.
To examine whether this also occurs during bar growth, we conducted a two-dimensional Fourier analysis following previous studies (details are provided in \ref{2DFA}).
The analysis shows that, even around the period of maximum bar-height growth, $\alpha_{20}$ tends to become more dominant as discharge decreases, and its magnitude is of the same order as $\alpha_{11}$.

Such dominance of $\alpha_{20}$ is likely attributable to the fact that, at lower discharge, the width-to-depth ratio increases and the hydraulic conditions approach the regime in which multiple-row bars can develop.
When considering mode reduction from multiple-row bars to alternate bars, it has actually been pointed out that accounting for interactions among multiple modes is important \cite{Izumi2002, Watanabe2006}.
When multiple modes with non-negligible amplitude are excited simultaneously, the available energy is not concentrated solely in the alternate-bar mode but is instead distributed among several modes.
As a result, even at the time when $\alpha_{11}$ reaches its equilibrium value, if other components are still growing, the growth rate of the composite bar height $\Omega_{H_{\mathrm{MB}}}$---which reflects the superposition of these components---becomes smaller than the growth rate of $\alpha_{11}$.
Although this hypothesis warrants further detailed testing, it may be useful to apply more general nonlinear analyses that account for intermodal interactions, even for alternate bars \cite{Fukuoka1985a, Fukuoka1985b, Kuroki1992}.
A more detailed explanation of this point, based on the theoretical analysis of \citeA{Kuroki1992}, is as follows.
When modal interaction with $\alpha_{20}$, two amplitude evolution equations are obtained. 
%%%%%%%%%%%%%%%%%%%%%%%%%%%%%%%%%%%
\begin{equation}
\frac{d\alpha_{11}}{dt}=C_1\alpha_{11}+C_2\alpha_{11}\alpha_{20},
\end{equation}
%%%%%%%%%%%%%%%%%%%%%%%%%%%%%%%%%%%
\begin{equation}
\frac{d\alpha_{20}}{dt}=C_3\alpha_{20}+C_4{\alpha_{11}}^2.
\end{equation}
%%%%%%%%%%%%%%%%%%%%%%%%%%%%%%%%%%%
Here $C_1,\ C_2,\ C_3,\ C_4$ denote coefficients estimated by nonlinear analysis.
The temporal evolution of bar height is then expressed as follows.
%%%%%%%%%%%%%%%%%%%%%%%%%%%%%%%%%%%
\begin{equation}
\frac{H_{\mathrm{MB}}}{h_0}=
\begin{cases}
\alpha_{11}+2\alpha_{20}+\dfrac{{\alpha_{11}}^2}{8\alpha_{20}} & (\alpha_{11}<4\alpha_{20}),\\
2\alpha_{11} & (\alpha_{11}>4\alpha_{20}).
\end{cases}
\end{equation}
%%%%%%%%%%%%%%%%%%%%%%%%%%%%%%%%%%%
Because this theoretical expression accounts for multiple modes, it is likely to reproduce the nonlinear evolution of bar height more accurately than Equation (\ref{Landau}).
It should be noted that $\Omega_{11}$ is defined for $m$ = 1 at the initially dominant wavelength, whereas $\Omega_{H_{\mathrm{MB}}}$ inherently includes the effect of wavelength elongation over time.
This is a limitation that cannot be accounted for within the framework of the existing theory.

%Another possible reason why $\Omega_{11}$ cannot adequately approximate $\Omega_{H_{\mathrm{MB}}}$ is that the theoretical analysis may not estimate the dominant wavenumber with sufficient accuracy.
%This could be examined if the time period during which an alternate-bar pattern appeared in the experiment could be identified, and one full wavelength at that stage could be obtained.
%However, in many cases, an initial multiple-bar pattern appeared first and then underwent mode reduction before an alternate-bar pattern formed, making it difficult to determine mechanically, without introducing author judgment, from which stage the bars should be regarded as alternate bars.
%In addition, because of the limitation in flume length, it was difficult to track one full bar wavelength over a long period.
%This is a limitation of the experimental method used in this study and remains an issue for future work.
%memo: 2dfftを時間ごとにおこなえば、$\alpha_{11}$の時間変化を定量化することができ、$\Omega_{11}$の実験値と理論値のより直接的な比較が可能となる。

\subsubsection{Applicability of Nonlinear Stability Analysis to Natural Rivers}
If a stability theory that addresses the issues identified in the preceding sections can be established, it is expected to provide a powerful tool for practical prediction of future changes in bar height.
In river management under climate change, hydrological variables such as air temperature and precipitation often receive the most attention.
From the perspective of river-engineering practice, it is also important to predict future bed evolution in response to such hydrological changes.
In recent years, nonlinear stability analysis has been applied to predict long-term variations in bar height \cite{Carlin2021, Redolfi2023, Inoue2024}.
Compared with numerical simulations that discretize and solve the governing equations, this approach has a much lower computational cost, making it well-suited for rapid evaluation across many scenarios.
The studies cited above, however, have not compared long-term discharge histories with the corresponding temporal evolution of bed morphology in natural rivers.
As a result, the applicability of the method remains insufficiently validated, and its limits across different river settings are still unclear.
Two steps are needed to advance field applications of nonlinear stability analysis.
The first is to examine the characteristics of the theoretical estimates through comparisons with laboratory experiments conducted under idealized conditions, as in this study.
Specifically, the findings presented in Sections \ref{result_exp} and \ref{lim} suggest that nonlinear stability analysis can potentially provide relatively accurate estimates of bar height in channels whose width-to-depth ratio is close to the critical value.
They also imply that the discharge dependence of the growth rate is strongly governed by the sediment-transport formula prescribed as an input to the analysis.
The second step is to assess site-specific validity through detailed comparisons between discharge histories and bed-morphology datasets in natural rivers.

\subsection{How Discharge Affects the Development of Bar Height?}
\label{mechanism}
Based on the experimental results in Section \ref{result_exp} and the hydraulic calculation results in Section \ref{result_cal}, we discuss how discharge influences the development of bar height.
Figures \ref{flow} and \ref{sediment} indicate that, after the flow is initiated over an initially flat bed, the spatial distributions of flow and sediment transport show little dependence on discharge while $H_\mathrm{MB}/d$ $<$  10.
In contrast, once $H_\mathrm{MB}/d$ exceeds approximately 10, differences in the spatial distributions of flow and sediment transport begin to emerge depending on the supplied discharge.
Under lower-discharge conditions, the flow depth is smaller; therefore, the bar height-to-depth ratio becomes relatively larger than under higher-discharge conditions.
Such conditions make the flow more susceptible to gravity-driven deflection, rather than deflection controlled solely by the bedform geometry itself (Figures \ref{contour}, \ref{flow}, and \ref{flow_shape}).
In addition, as depositional areas develop, the flow becomes weaker immediately downstream of these areas and sediment transport ceases there (Figures \ref{contour} and \ref{sediment}).
This reduces the mobility of the bar position, or equivalently, the migration of the scour zone (Figures \ref{speed} and \ref{height_vs_x}).
A similar stabilization mechanism, involving shallow flow and suppressed bedload transport over bar surfaces, has been reported for stationary alternate bars in a steep channel with mixed-size sediment \cite{Lisle1991}.
During this stage, the depositional areas expand in areal extent (Figure \ref{zerocross}), but the bed elevation does not change substantially (Figure \ref{dpst_scr}).
Flow deflection, the expansion of non-sediment-transport zones, and the weakening of bar migration tend to confine flow and sediment transport to the scour zones of the bars, thereby enhancing local scour.
This creates conditions favorable for the development of bar height.
However, as suggested by Figures \ref{flow}, \ref{sediment}, and \ref{vs_qsfml}, bar-height growth can further enhance flow deflection, promote the expansion of non-sediment-transport zones, and suppress bar migration.
These processes should therefore not be interpreted as a simple one-way causal relationship; rather, they are likely to interact with one another in a way that mutually reinforces their effects.
Conversely, under high-discharge conditions, the processes described above are less pronounced, and bar-height growth is therefore considered to be relatively suppressed.

This mechanism may also explain the previous experimental finding of \citeA{Redolfi2020} that, under fully wet conditions in which bars remain submerged, bar height becomes larger at lower discharge.
Although the sidewalls were fixed in the present study, if erodible banks were used, bank erosion would be expected to progress together with the growth of bar height.
\citeA{Visconti2010} experimentally showed that bank erosion progresses when discharge is reduced.
They referred to this behavior as a “triggering process”, and the mechanism described above may provide a physical explanation for this process.
Taken together, these results provide quantitative support for previous qualitative interpretations of how bar height varies with discharge.
\section{Conclusions}
In this study, we quantified the dependence of the growth rate of alternate-bar height on water discharge and developed a mechanistic explanation for this behavior. 
These advances were made possible by combining high-frequency measurements of bed topography in flume experiments, coefficient estimation for the Landau equation, and hydraulic calculations performed over the measured bed topography. 

We quantified the temporal evolution of bar height using Stream Tomography, which measures bed topography without interrupting the flow.
We found that the growth of bar height was driven primarily by progressive scour, not by deposition.
We also demonstrated that the growth rate and equilibrium bar height can be robustly identified for each condition by fitting the general solution of the Landau equation to the resulting bar-height time series.
The inferred dimensionless growth rate of bar height shows a general tendency to increase with decreasing discharge, but its sensitivity to discharge is weaker than predicted by linear stability analysis, and the discrepancy increases as discharge decreases.
This discrepancy may reflect the breakdown of the linearization assumptions in the analysis, the non-negligible interactions among multiple modes, and the elongation of wavelength in the experimental system.
These ST-based growth rate estimates provide foundational data for testing the validity and limits of stability theory and may help improve predictions of future bar-height changes in natural rivers.

We further interpreted how the bar growth rate varies with discharge using fixed-bed hydraulic calculations over ST-derived bed topography, flow deflection and the extent of non-sediment-transport zones.
We quantitatively showed that lower-discharge conditions create hydraulic conditions favorable for bar development.
Specifically, during the stage of bar development, flow deflection becomes more pronounced under the influence of gravity, while sediment transport is restricted immediately downstream of the depositional area, reducing bar mobility.
These processes concentrate flow and sediment transport in the scour zones, thereby enhancing local scour and promoting the growth of bar height.
These findings deepen our understanding of the mechanisms governing bar development.
%========================================================6============================================================

\section*{Conflict of Interest}
The authors declare no conflicts of interest relevant to this study.

\section*{Data Availability Statement}
The experimental and numerical data that support the findings of this study are available in the dataset described by \citeA{Dataset}.

\acknowledgments
The work was supported by JSPS KAKENHI (Grant Nos. JP24H00365 and JP24KJ1171), and JST CREST (Grant No. JPMJCR25Q5). 
During manuscript preparation, generative AI tools (ChatGPT and Claude) were used to check language grammar, phrasing, and clarity.
The authors reviewed and edited all AI-assisted text and take full responsibility for the final manuscript.

\bibliography{ref}

\newpage
\appendix

\section{Supplementary Information on the Experimental Results}
This section provides information supporting the accuracy of the experimental results presented in Section \ref{result_exp}.

\subsection{Landau-Equation Fit}
\label{LEF}
\begin{figure}[t]
 \centering
   \includegraphics[width=0.6\textwidth]{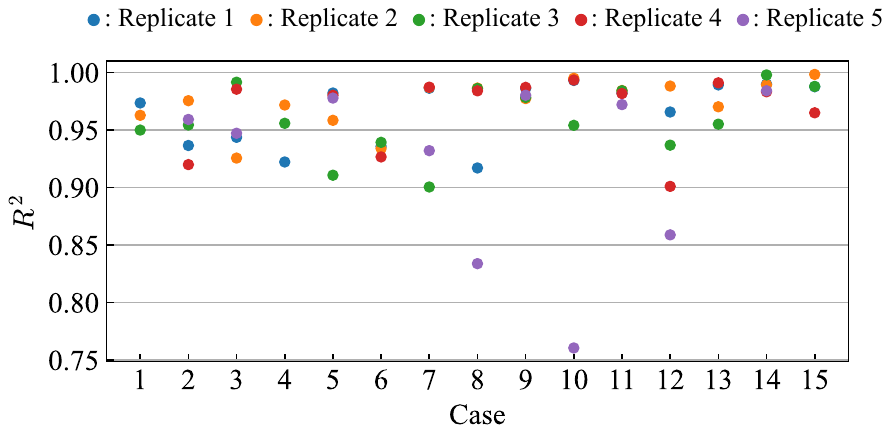}
 \caption{
 Accuracy of the Landau-equation fit. 
 The coefficient of determination $R^2$ exceeds 0.90 for most cases, and it is approximately 0.75 even for the poorest-fitting case.
 This indicates that the Landau-equation fit is adequate.
 }
 \label{r_squared}
\end{figure}

This section describes in detail the fitting accuracy of the Landau equation used in Section \ref{appl_LE}.
Figure \ref{r_squared} shows, for each condition, the coefficient of determination $R^2$, which quantifies the goodness of fit of Equation (\ref{solution}) to the measurements. 
$R^2$ exceeds 0.90 for most conditions and is still on the order of 0.75 even for the least favorable case. 
These results indicate that it is reasonable to represent the temporal evolution of bar height using the functional form of the Landau equation.
The statistical reliability of the fitted coefficients was assessed by hypothesis testing, and any coefficients with $p$-values greater than 0.05 were excluded from the analysis.
This procedure also enables statistical identification of conditions in which the bar height did not reach equilibrium.
Even after this screening, at least three replicates remained for each condition.

\subsection{Equilibrium Bar Height}
\label{EBH}
Here we examine whether the bar heights obtained in our experiments are reasonable compared with previous studies.
To this end, we present the equilibrium values among the coefficients obtained from fitting the Landau equation in Section \ref{appl_LE}.
The empirical relationship proposed by \citeA{Ikeda1984}, given below, is shown as a dashed line in the figure for comparison.
%%%%%%%%%%%%%%%%%%%%%%%%%%%%%%%%%%%
\begin{equation}
\frac{H_\mathrm{Be}}{h_0}=
\left(\frac{B}{d}\right)^{-0.45}
9.34\exp
\left(
2.53\ \mathrm{erf}
\frac{\log_{10}\dfrac{B}{h_0}-1.22}{0.594}
\right).
\label{Heq_Ikeda}
\end{equation}
%%%%%%%%%%%%%%%%%%%%%%%%%%%%%%%%%%%
In the above equation, the bar height, $H_\mathrm{B}$, defined by \citeA{Ikeda1984} is the difference between the maximum and minimum bed elevations on the cross section that contains the minimum bed elevation within one bar wavelength.
$H_\mathrm{Be}$ denotes its equilibrium value.

\begin{figure}[t]
 \centering
   \includegraphics[width=1.0\textwidth]{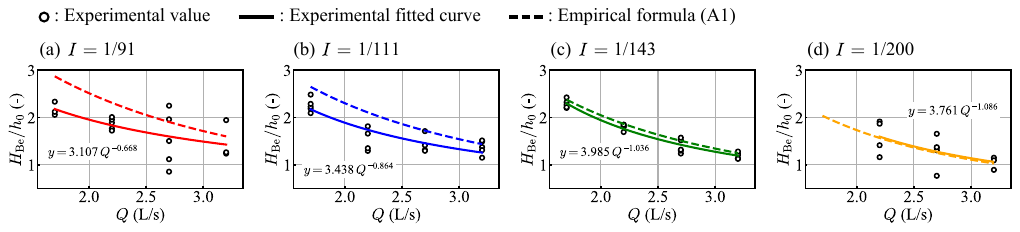}
 \caption{
 Relationship between the equilibrium bar height and discharge at each channel slope.
 The empirical relationships (\ref{Heq_Ikeda}) in this range increase monotonically with decreasing discharge, so we fitted the relationship between the experimental values $y$ and discharge using a simple monotonic power-law function $y$ = $b$ $\cdot$ $Q^a$, and plotted the regression as a solid line.
 The sensitivity to discharge can be assessed from the value of $a$.
 The equilibrium bar heights obtained in our experiments broadly agree with empirical relationships from previous studies in both order of magnitude and dependence on discharge.
 }
 \label{heq}
\end{figure}

Figure \ref{heq} shows that the measured equilibrium bar height increases monotonically with decreasing discharge, consistent with the empirical relationship.
The reported accuracy of Equation (\ref{Heq_Ikeda}) is +85\ \% $\sim$ -45\ \% \cite{Ikeda1984}.
Within this uncertainty range, the experimental values obtained in this study are considered reasonable.
Note that the cases with $I$ = 1/143 show particularly small inter-replicate variability and a high degree of agreement between the measurements and the empirical relationship.

\subsection{Sediment Transport Formula}
\label{STF}
\begin{figure}[t]
 \centering
   \includegraphics[width=1.0\textwidth]{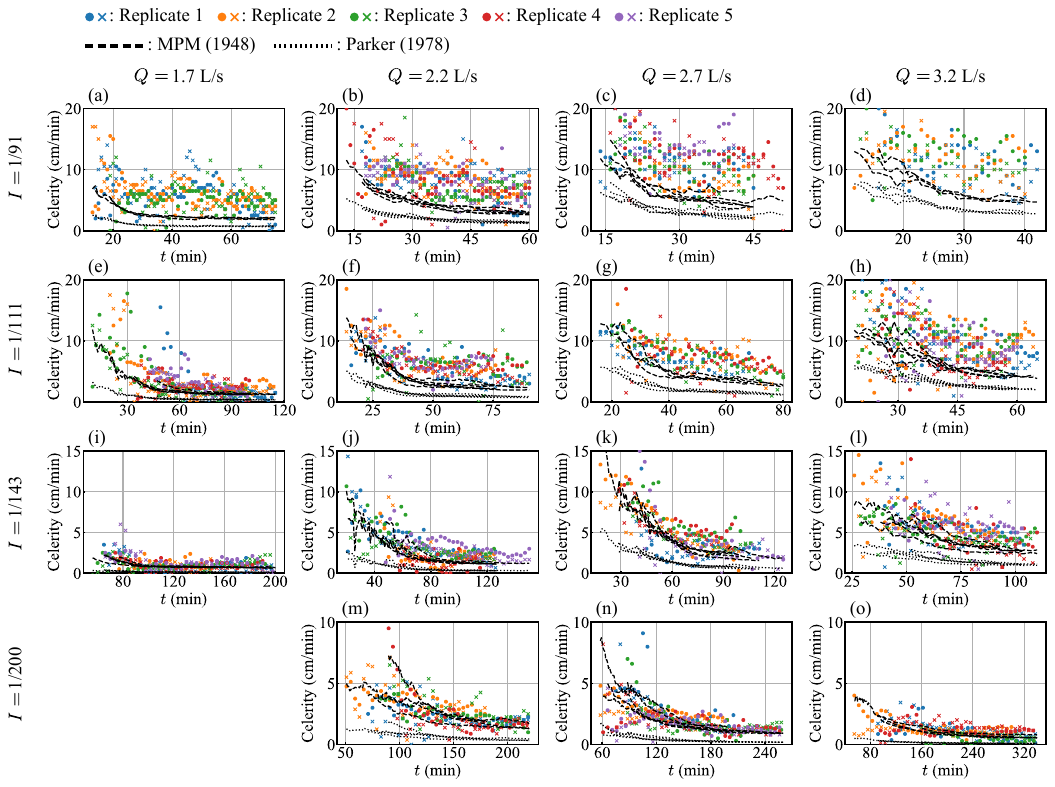}
 \caption{
 Temporal changes in the measured bar celerity and the estimated bar celerity obtained using the \citeA{mpm} (dashed line) and \citeA{Parker1978} (dotted line) sediment transport formulas.
 Rows correspond to channel slope ($I$ = 1/91, 1/111, 1/143, and 1/200, from top to bottom), and columns correspond to discharge ($Q$ = 1.7, 2.2, 2.7, and 3.2\ L/s, from left to right).
 Circles and crosses indicate the upstream and downstream ends of the target bar, respectively.
 For convenience in estimating bar celerity, the values were subsampled from the full dataset.
 Under all conditions, the bar celerity estimated using the \citeA{mpm} formula shows better agreement with the measured values than that estimated using the \citeA{Parker1978} formula.
 }
 \label{vs_qsfml}
\end{figure}

In Section \ref{result_exp}, we examine the sediment transport formula most appropriate for the present experiments in order to scale experimental time and perform the theoretical analysis.
However, even in laboratory flumes, it is difficult to measure sediment transport accurately in both space and time.
We therefore estimate sediment transport indirectly.
Specifically, we make use of its relationship with the measured bar celerity and bar height.
According to \citeA{Fujita1975}, bar celerity $C$, bar height $Z$, and sediment transport are related as follows.
%%%%%%%%%%%%%%%%%%%%%%%%%%%%%%%%%%%
\begin{equation}
C = \frac{q_{\mathrm{s}}}{2(1-p)Z}.
\label{Fujita_fml}
\end{equation}
%%%%%%%%%%%%%%%%%%%%%%%%%%%%%%%%%%%
In the study of \citeA{Fujita1975}, $Z$ was probably defined as the spatially averaged bar height. In the present study, however, because we focus on an individual bar, we assume that $Z$ $\approx$ $H_\mathrm{MB}$.
For the reasons described in the main text, we compared the \citeA{mpm} formula and the \citeA{Parker1978} formula.
Figure \ref{vs_qsfml} shows the temporal changes in the measured bar celerity and the corresponding values given by Formula (\ref{Fujita_fml}).
The figure shows that, under all conditions, the celerity formula based on the \citeA{mpm} formula reproduces the observed bar celerity better than that based on the \citeA{Parker1978} formula.
Although the celerity formula based on the \citeA{mpm} formula differs from the observations by several cm/min under the $I$ = 1/91 condition, it agrees with the observations to within 1 cm/min under the $I$ = 1/143 and $I$ = 1/200 conditions.

\section{Two-Dimensional Fourier Analysis}
\label{2DFA}
\begin{figure}[t]
 \centering
   \includegraphics[width=1.0\textwidth]{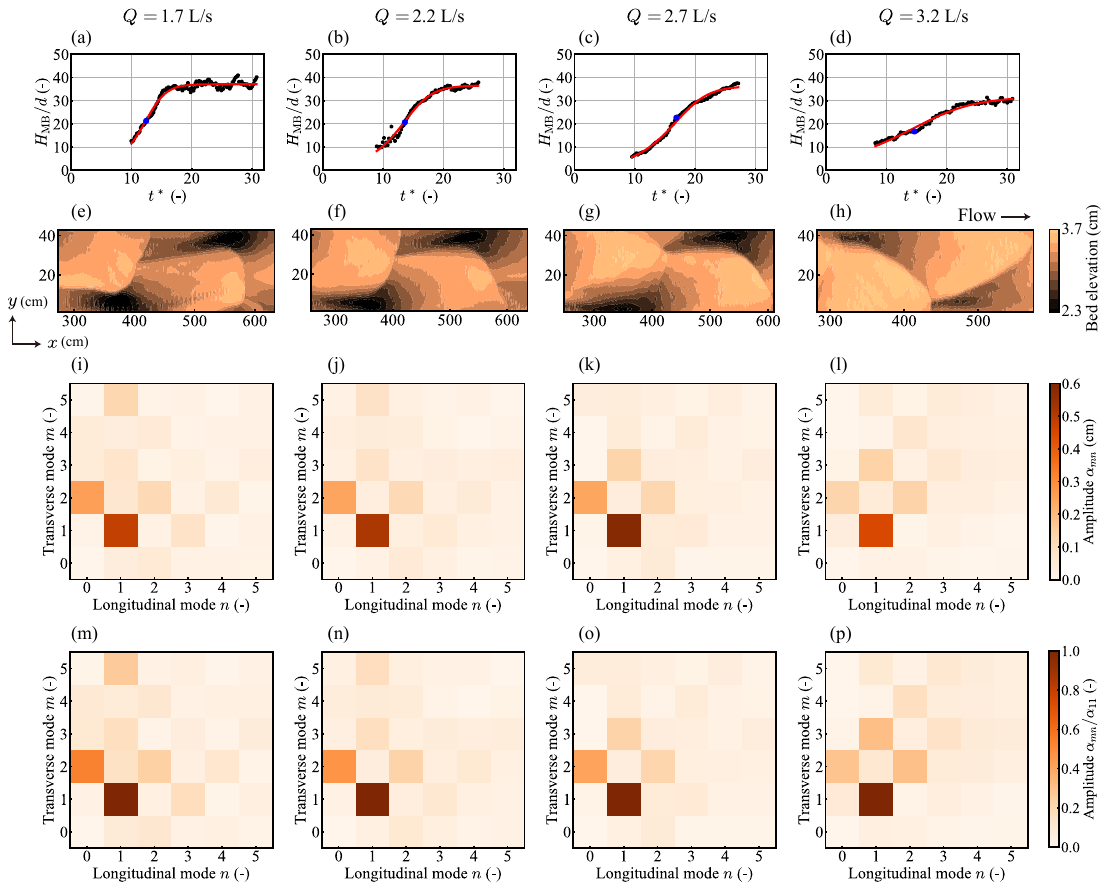}
 \caption{
 First row, panels (a)--(d): Temporal evolution of bar height, as in Figure \ref{Landau_fit}, showing the identification of the target time. The blue circles indicate the time at which the growth speed is largest.
 Second row, panels (e)--(h): Bar morphology used for the analysis.
 Third row, panels (i)--(l): Amplitude of each wavenumber component, $\alpha_{mn}$.
 Fourth row, panels (m)--(p): $\alpha_{mn}$ normalized by $\alpha_{11}$.
 Columns correspond to discharge ($Q$ = 1.7, 2.2, 2.7, and 3.2\ L/s, from left to right).
 Although $\alpha_{11}$ is the most dominant component, other components are also of comparable magnitude.
 }
 \label{Fourier}
\end{figure}

This section provides details of the two-dimensional Fourier analysis described in Section \ref{multi-mode}.
Figure \ref{Fourier} shows, as an example, the results for each discharge condition for $I$ = 1/143.
From the time series of bar height shown in the first row, the time at which its time derivative reaches a maximum is identified, and the bar morphology at that time (second row) is used as the analysis target.
The third row shows the amplitude for each wavenumber $\alpha_{mn}$.
The fourth row also shows $\alpha_{mn}$ normalized by $\alpha_{11}$.
The two-dimensional fast Fourier transform was performed using the NumPy function \texttt{numpy.fft.fft2} in Python.
The figure indicates that $\alpha_{11}$ is the most dominant component for all discharge conditions, while other components are also appreciable.

\begin{figure}[t]
 \centering
   \includegraphics[width=0.8\textwidth]{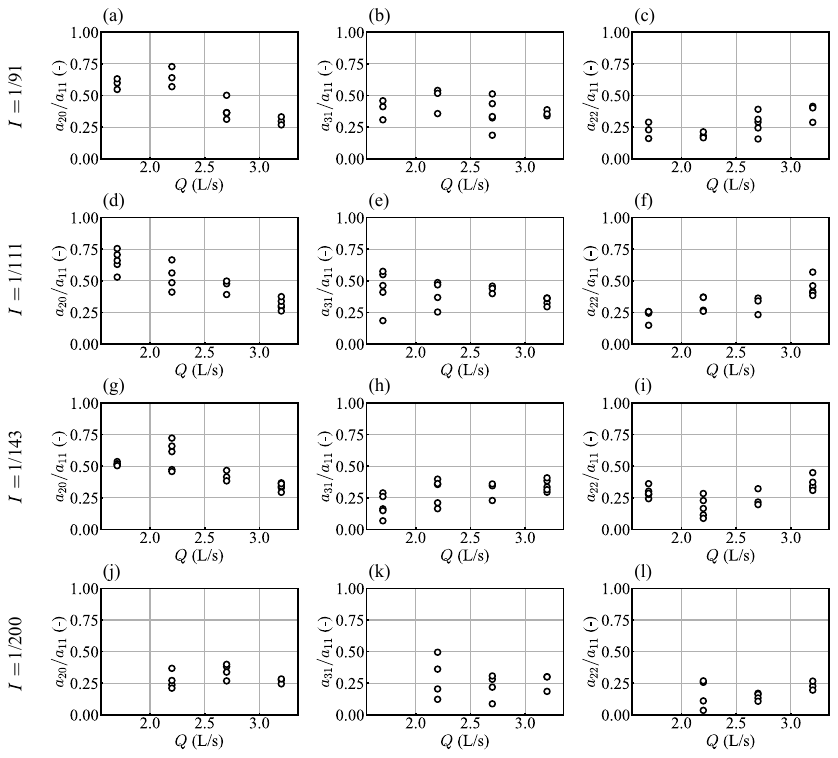}
 \caption{
 Relationship between discharge and the relative amplitude of each Fourier component, $\alpha_{mn}$, relative to the primary alternate-bar mode $\alpha_{11}$.
 First column, panels (a), (d), (g), and (j): $\alpha_{20}/\alpha_{11}$.
 Second column, panels (b), (e), (h), and (k): $\alpha_{31}/\alpha_{11}$.
 Third column, panels (c), (f), (i), and (l): $\alpha_{22}/\alpha_{11}$.
 Rows correspond to channel slope ($I$ = 1/91, 1/111, 1/143, and 1/200, from top to bottom).
 Replicates for which the alternate-bar pattern was not clearly identifiable, and for which wavenumber analysis was therefore considered difficult, were excluded.
 The dominance of $\alpha_{20}$ becomes more pronounced under lower-discharge conditions, and its magnitude is of the same order as that of $\alpha_{11}$.
 }
 \label{Fourier_vsQ}
\end{figure}

Figure \ref{Fourier_vsQ} summarizes the relationship between discharge and each Fourier component for all channel slopes considered in this study.
We examine three components commonly regarded as representative wavenumber components of alternate bars, $\alpha_{20}$, $\alpha_{31}$, and $\alpha_{22}$ \cite{Hasegawa1982, Garcia1993, Redolfi2020}.
Components are presented as ratios to $\alpha_{11}$.
The figure shows that, for all channel slopes, $\alpha_{31}$ and $\alpha_{22}$ exhibit no clear dependence on discharge, whereas $\alpha_{20}$ tends to become more dominant as discharge decreases.
The ratio $\alpha_{20}/\alpha_{11}$ reaches up to approximately 0.8, suggesting that it is not negligible for the development of alternate bars.

\end{document}